\documentclass[11pt]{article}

\usepackage{physics} 
\usepackage{siunitx} 
\usepackage{enumerate} 
\usepackage{pgfplots}
\usepackage{pgfplotstable}
\usepackage{tikz,pgfplots}
\usepackage{cite}
\usepackage{amsmath} 
\usepackage{wasysym} 
\usepackage{amsfonts}
\usepackage{multirow}
\usepackage{float} 
\usepackage{placeins} 

\usepackage{longtable}
\usepackage{hyperref}
\usepackage{float}
\usepackage{hyperref}
\usepackage{float}
\usepackage{lipsum}
\usepackage{authblk}
\usepackage{soul} 

\usepackage{subcaption}
\usepackage[square,numbers,sort&compress]{natbib}
\usepackage{amsmath,amssymb}

\usepackage{geometry}
\begin{document}
	\title{\bf The imitation game reloaded:  effective shadows \\ of dynamically robust spinning Proca stars} 

\author[1]{Ivo Sengo\thanks{sengo@ua.pt}}
\author[1]{Pedro V.P. Cunha\thanks{pvcunha@ua.pt}}
\author[1]{ Carlos A. R. Herdeiro\thanks{herdeiro@ua.pt}}
\author[1]{Eugen Radu\thanks{eugen.radu@ua.pt}}
\affil[1]{Departamento de Matemática da Universidade de Aveiro and Centre for Research and Development
	in Mathematics and Applications (CIDMA), Campus de Santiago, 3810-183 Aveiro, Portugal}
	
\date{February 2024}

\maketitle
\begin{abstract}
We analyse the lensing images by dynamically robust rotating (mini-)Proca stars surrounded by thin accretion disks. Due to their peculiar geodesic structure we show that these images exhibit striking similarities with the ones of BHs, for appropriately chosen disk intensity profile, when imposing a GRMHD-motivated emission cut off. Additionally, and unlike the non-rotating case, these similarities prevail even when considering equatorial observations. This example illustrates how a horizonless compact object without light rings, with a plausible formation mechanism and dynamically robust, could mimic detailed features of black hole imagiology.
\end{abstract}

\newpage
\tableofcontents

\section{Introduction}
\label{sec:Intro}

Once imperceptible to our detectors, black holes (BHs) candidates are increasingly becoming standard science targets for precise tests of General Relativity (GR)~\cite{EventHorizonTelescope:2019dse,LIGOScientific:2016aoc,GRAVITY:2018ofz}. 

Although BHs have been theoretically well established for over 50 years, our understanding of BH physics still faces important limitations, notably the formation of singularities~\cite{Penrose:1964wq} within the event horizon and the puzzle of information loss~\cite{Hawking:1976ra}. These limitations motivate the exploration of alternative (and sometimes more exotic) classes of gravitational entities. One such alternative comes in the form of compact horizonless compact objects, offering a fresh perspective beyond the traditional BH paradigm~\cite{Cardoso:2019rvt}.

 These speculative horizonless entities, which are regular objects without an horizon and are devoid of singularities, offer a promising framework for modelling very compact gravitational objects with potential astrophysical viability~\footnote{Here, ``astrophysically viable solutions'' refers to solutions that exhibit dynamical stability, possess a formation mechanism, and are well motivated by beyond standard model theories or well posed GR alternatives.} that could mimic a BH. In particular, could the BH candidates observed by the Event Horizon Telescope (EHT) collaboration actually be consistent with such an horizonless compact object (HCO)~\cite{EventHorizonTelescope:2022xqj}?

To replicate the optical appearance of a BH, a HCO must first accurately mimic the most distinctive feature of a black hole image: its shadow~\cite{Falcke:1999pj,Cunha:2018acu}. The latter arises from the absorption of light that comes sufficiently close to the trapped region and its appearence reflects the extreme gravitational bending of light around the BH, manifesting itself as a dark, silhouette-like region contrasting against background light. 
In the literature, different examples of HCO with shadow like features (effective shadows) can be found, most of them falling in one of the following categories~\footnote{These are not mutually exclusive, as some models may combine features that fall in more than one category (e.g., the work discussed in \cite{Rosa:2022tfv})}:

\begin{enumerate}	
	\item \textbf{Re-emission with energy loss}. It has been observed that some spacetimes exhibiting highly chaotic null geodesics~\cite{Cunha:2015yba,Cunha:2016bjh,Bacchini:2021fig,Bah:2023ows}, light can become confined within the central gravitational well of an object for an extended time. If the HCO is composed of a micro-structure that locally couples with photons, then radiation may gradually lose energy during interactions with these structures. Ultimately, photons that escape the trapped zone would be carrying negligible energy, thereby creating an apparent shadow~\cite{Bacchini:2021fig}. It is so far unclear if such trapped regions, akin to light rings \cite{Cunha:2022gde}, could potentially induce spacetime instabilities.

	\item \textbf{Disk red-shift}.  In cases where the HCO attains large compactness, the red-shift effects can increase significantly. Consequently, when examining a thin disk model that extends far within the central object, the luminosity emitted from within is markedly dimmer in comparison to the outer regions, creating an shadow-like feature in the image~\cite{Rosa:2022tfv}.
	\item  \textbf{Disk cutoff}. An effective shadow can emerge when an accretion disk exhibits a cutoff in its profile~\cite{Herdeiro:2021lwl}. A possible disk cutoff might be associated with characteristics of the geodesic structure, such as the existence of an Innermost Stable Circular Orbit (ISCO). This disk cutoff might may be further motivated from the accretion flow dynamics, for instance the stalling of accretion due to the inneficiency of magneto-rotational  (MRI) instabilities~\cite{Olivares:2018abq}.
 
\end{enumerate}

In this work we explore a family of examples falling within the third category. Unlike the previous examples in~\cite{Olivares:2018abq,Herdeiro:2021lwl}, however, the family of backgrounds explored here are dynamically robust. Indeed, in~\cite{Olivares:2018abq} the case study with an effective shadow was an unstable scalar boson star, whereas the case study in~\cite{Herdeiro:2021lwl} were static Proca stars (PSs) recently found to be also unstable~\cite{Herdeiro:2023wqf}.   
Herein, we will be exploring rotating solutions within the Proca model's fundamental branch~\cite{Santos:2020pmh}. These solutions, some of which are dynamically stable, also emerge as intriguing candidates potentially sourcing observed gravitational wave events \cite{CalderonBustillo:2020fyi,CalderonBustillo:2022cja}.

In the study by Olivares et al.~\cite{Olivares:2018abq}, cutting-edge General Relativistic Magnetohydrodynamics (GRMHD) simulations were employed in the background of specific scalar boson star models, assessing their potential ability to mimic the appearance of a BH. While the outcomes revealed that the considered boson stars did not perfectly mimic black holes, some configurations did produce an apparent shadow. This shadow arose due to a stalled accretion flow, which the authors associated with the presence of a maximum in the angular velocity profile of timelike circular orbits, which suppressed the Magneto-Rotational Instability (MRI) within that region.

The observed size of the effective shadow, as documented in~\cite{Olivares:2018abq}, appeared notably smaller in comparison to that of a Schwarzschild  BH under equivalent conditions. Building upon the findings of~\cite{Olivares:2018abq}, the work~\cite{Herdeiro:2021lwl} demonstrated that by considering spherical PS solutions, it became feasible to achieve an effective shadow more closely resembling that of a Schwarzschild BH, under similar observational conditions. However, despite obtaining a striking similarity between a BH and a boson star when observed face-on ($\theta=0$), this resemblance was less pronounced for equatorial observations. This discrepancy stemmed from the fact that these stars were not compact enough to induce strong gravitational lensing~\footnote{Namely the wide ``envelope'' created by the lensed images of the accretion disk area behind the compact object (see \cite{James:2015yla})}.

In a subsequent development, the spherical Proca solutions considered in~\cite{Herdeiro:2021lwl} were found to be unstable: they decay into a more fundamental static state of the Proca family, which curiously are geometrically non-spherical~\cite{Herdeiro:2023wqf}. This instability undermines the astrophysical viability of the Proca solutions considered in~\cite{Herdeiro:2021lwl}.

In the present work we aim to reconsider the Proca family as viable BH mimickers, when lit by a geometrically thin and optically thick accretion disks. Building upon the results of \cite{Olivares:2018abq,Herdeiro:2021lwl}, and capitalizing on the intricate geodesic structure governing timelike circular orbits, we showcase how specific rotating PS configurations can play the BH imitation game in a better way than the spherical case. We emphasise that most solutions considered here are free of bound photon orbits $i.e.$ light rings, thus avoiding the light ring instability discussed in~\cite{Cunha:2022gde}, which generically plagues ultracompact HCO with a plausible formation mechanism~\cite{Cunha:2017qtt}.

This paper is organized as follows. Section~\ref{sec:TCO} provides an overview on the geodesic structure of timelike circular orbits around generic compact objects. In section~\ref{sec:PSs} we review the solution space of rotating PSs, together with some properties of a few selected solutions. In section~\ref{sec:Disk} we discuss the accretion disk profile for these stars. In sections ~\ref{sec:lensing}$-$\ref{sec:redshift} we present the gravitational lensing images of these solutions. Finally, in section~\ref{sec:conclusions} we offer some conclusions.

\section{TCO for generic compact objects -- overview }
\label{sec:TCO}

The behaviour of an accretion disk encircling a compact object is influenced not solely by the spacetime geometry around the central object, but also by a variety of surrounding environmental factors~\cite{Abramowicz:2011xu}. Nevertheless, one can determine the potential locations of these disks by examining the arrangement of timelike circular orbits (TCOs) within a specific spacetime geometry.

As a first approximation, we can conceptualize accretion disks as a continuous series of stable circular paths followed by timelike geodesics, situated within the equatorial plane of the compact object. In the case of Kerr geometries, these stable TCOs extend from spatial infinity down to a radial coordinate marked as $r_{\mathrm{ISCO}}$, representing the innermost stable circular orbit (ISCO). Below this value of $r_{\mathrm{ISCO}}$, stable TCOs cease to exist.

As highlighted in~\cite{Delgado:2021jxd}, for generic non-Kerr compact objects, the arrangement of TCOs can be more intricate: unlike the Kerr scenario, there can exist solutions featuring multiple disconnected regions where stable TCOs are possible. These distinct ``stable regions'' might be delimited by zones where only unstable TCOs are allowed or by areas where TCOs do not exist altogether. These scenarios introduce novel possibilities for accretion disk configurations, potentially leading to new gravitational lensing effects.

\subsection{Determining the structure of TCOs}

For the purposes of this analysis, we shall be interested in finding the regions where stable equatorial TCOs are possible on a given  stationary, axi-symmetric, circular, asymptotically flat, 1+3 dimensional spacetime -- these are the regions where we shall place the accretion disks. This problem was recently addressed in \cite{Delgado:2021jxd}, and below we review its main conclusions.

For the aforementioned assumptions, we can describe a generic (circular) spacetime with the following metric\footnote{With the signature $\left(-,+,+,+ \right)$}.

\begin{align}
	ds^2 = g_{tt} dt^2 +2 g_{t\phi} dt d\phi + g_{\phi \phi} d \phi^2 +  g_{rr} dr^2 + g_{\theta \theta} d\theta^2 \, . 
 \end{align}
 
 Following \cite{Delgado:2021jxd}, we shall introduce the following definitions:
 
 \begin{align}
 	& \mathcal{B}(r, \theta) \equiv g_{t \phi}^2 -g_{tt} g_{\phi \phi} >0 \, ,\\
 	& \mathcal{C}(r, \theta) \equiv g_{t \phi}'^2 - g_{tt}' g_{\phi \phi}' \, , \\
 \end{align}
 
\noindent where the prime denotes the derivative w.r.t the radial coordinate. Moreover, the integrals of motion associated with the Killing vectors allow us to write the energy, $\varepsilon$, and the angular momentum, $\ell$, as

\begin{align}
	-&\varepsilon \equiv g_{tt} \dot{t} + g_{t \phi} \dot{\phi} \, , \\
	& \ell \equiv g_{t \phi} \dot{t} +g_{\phi \phi} \dot{\phi} \, .	
\end{align}

We also introduce the effective potential $V(r)$ \footnote{In reference \cite{Delgado:2021jxd}, this corresponds to the potential $V_{-1}(r)$; since we will only be interested in timelike orbits, we have dropped the subscript for the sake of clarity and simplicity. }:
 \begin{align}
 	V(r)=-1 +\frac{\mathcal{A}(r,\varepsilon,\ell)}{\mathcal{B}(r)} \, , 
 \end{align}
 
\noindent where $\mathcal{A}(r, \varepsilon, \ell) \equiv g_{\phi \phi} \varepsilon^2 + 2g_{t\phi} \varepsilon \ell + g_{tt} \ell^2$.

If circular orbits exist, their angular velocity (as measured at infinity) can be computed as 

\begin{align}
	\Omega_{\pm} = \frac{-g'_{t \phi} \pm \sqrt{\mathcal{C}(r)}}{g'_{\phi \phi}} \, . 
\end{align} 

Throughout the article, we shall use the subscript plus (minus) to represent co-rotating (counter-rotating) orbits.

\subsubsection*{Scheme for determination of the allowed regions}
If stable TCOs exist for a given coordinate $r*$, then the following conditions must be satisfied:

\textbf{1) Real energy and angular momentum}. Circular geodesics on $r*$ must be endowed with real energy and angular moment. This is the case \textit{iff}:

\begin{align}
\beta_{\pm} \equiv  \left( -g_{tt} -2g_{t\phi} \Omega_{\pm} -g_{\phi \phi} \Omega^2_{\pm}   \right)\Bigr|_{\substack{r*}}  = -\mathcal{A}(r*, \Omega_{\pm}, 1) >0 \, .
\end{align}

If the latter is not verified, then at the coordinate $r*$ one would instead find spacelike circular orbits, which do not represent the motion of physical particles.\\

\noindent  \textbf{2) Real angular velocity}. The angular velocity measured at infinity must be real. This is guaranteed if $\mathcal{C}(r*)\geq 0$.\\

\noindent  \textbf{3) Stability}. If stable TCO exist, then the second derivative of the effective potential $V(r)$ must satisfy $V''(r*)>0$.

\section{TCOs structure for Proca stars}
\label{sec:PSs}

We will focus on PSs, the horizonless solutions of the Einstein-complex-Proca model
 \begin{align} 
S = \int \mathrm{d}^4 x \sqrt{-g} \left( \frac{R}{16 \pi}  - \frac{1}{4} F_{\alpha \beta} \bar{F}^{\alpha \beta} - \frac{1}{2} \mu^2 A_{\alpha} \bar{A}^\alpha \right) \, , 
\end{align}

\noindent where $g$ is the metric determinant, $R$ is the Ricci scalar, $\mu$ is the vector field mass and $F_{\mu \nu} = \partial_{\mu}A_{\nu} -\partial_{\nu}A_{\mu} $ is the field strength written in terms of the 4-potential $A_{\mu}$. The Proca potential ansatz has the form

\begin{align}
\mathbf{A}= \mathrm{e}^{\mathrm{i}\left( m \phi - \omega t \right)} \left( \mathrm{i}V dt +H_1 dr + H_2 d \theta + \mathrm{i}H_3 \sin \theta d \phi \right) \, ,
\end{align}

\noindent where the four functions $\left( V, H_i \right)$ all depend on $\left( r, \theta \right)$.  This ansatz has an harmonic time and azimuthal dependence, associated with the frequency $\omega > 0 $ and the azimuthal harmonic index $m \in \mathbb{Z} $, respectively.

\begin{figure}[H]
	\centering
	\includegraphics[width=0.6\columnwidth]{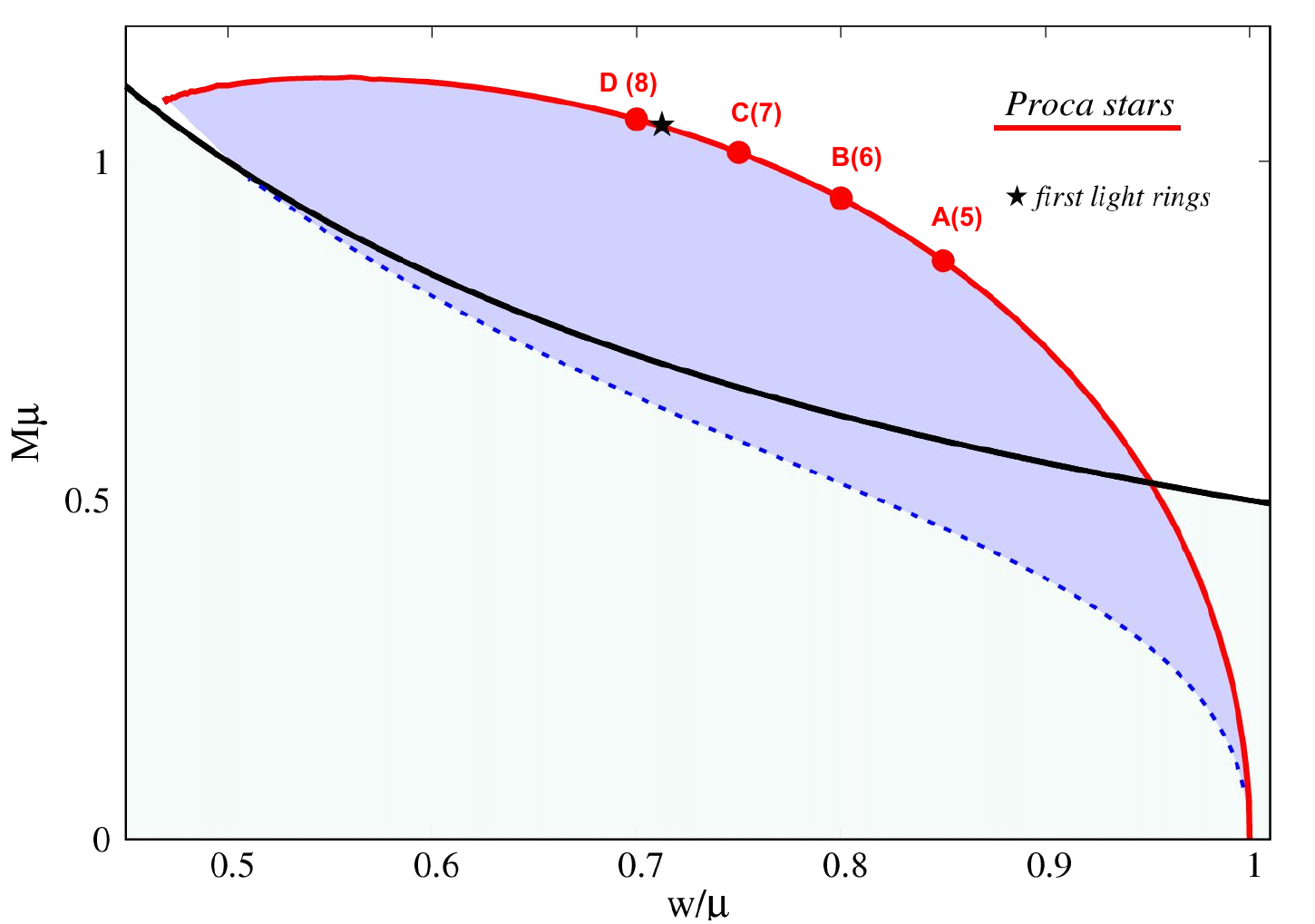}
	\caption{\small \label{fig:solution-space} The domain of existence of PSs (red solid line).  The  star symbol on the PS line marks the first PS solution to feature a light ring (LR) orbit. We have highlighted the four solutions that we are going to focus on  throughout the paper. These solutions are going to identified as A, B, C and D. The value in parenthesis correspond to the labelling used in our previous work \cite{Sengo:2022jif}. The shaded blue region correspond to the existence domain of BHs with synchronised Proca hair, which interpolate PSs to the Kerr existence line (dotted blue line). The reader is referenced to~\cite{Herdeiro:2016tmi,Sengo:2022jif} for further details.}
\end{figure}

We will be interested  on the fundamental solutions of this model ($m=1$) with no radial nodes of $V$. The construction and properties of these solutions where studied in~\cite{Herdeiro:2016tmi,Santos:2020pmh}, whereas the lensing images have been recently explored in~\cite{Sengo:2022jif}. In Fig.~\ref{fig:solution-space} we show the domain of existence of these solutions (red solid line) displayed in a $(M\mu\,,\,\omega/\mu)$ diagram, where $M$ is the total mass of the solution. In Fig.~\ref{fig:solution-space} we have highlighted four illustrative solutions which will be the focus throughout this paper. These solutions are going to be identified as $A$, $B$, $C$ and $D$, respectively labelled with the numbers \{5,6,7,8\} in the previous work~\cite{Sengo:2022jif}.

The four solutions $A$, $B$, $C$, and $D$ represent distinct categories of TCO structures. Case $A$ denotes a solution featuring stable TCOs across the entire spatial domain, encompassing both co-rotating and counter-rotating orbits. In case $B$, the co-rotating orbits remain stable throughout the space, whereas an unstable region emerges for the counter-rotating orbits in the radial range $r\in [r_1,r_2]$. Similarly, case $C$ exhibits stable co-rotating orbits, with a yet larger unstable region for the counter-rotating ones. Notably, this region progressively expands further as one moves along the PS existence line in the direction $A\to D$. Case $D$ is representative of solutions characterized by a pair of Light Rings (LRs), with a stable LR at the radial coordinate $r_{LR1}=r_1$, whereas the unstable LR lies between $r_1$ and $r_2$, $i.e.$ $r_1<r_{LR2}<r_2$. As elaborated in~\cite{Delgado:2021jxd}, TCOs are prohibited in the range $r\in [r_{LR1}, r_{LR2}]$ between these two LRs. These distinct scenarios are visually depicted in Fig.~\ref{fig:TCO-structure}.\\ 

In the next section we discuss the placement of the disk with respect to the different TCO regions that we have presented here.

\begin{figure}[H]
	\begin{tabular}{cc}
		\subfloat[Case A]{\includegraphics[width=0.45\columnwidth]{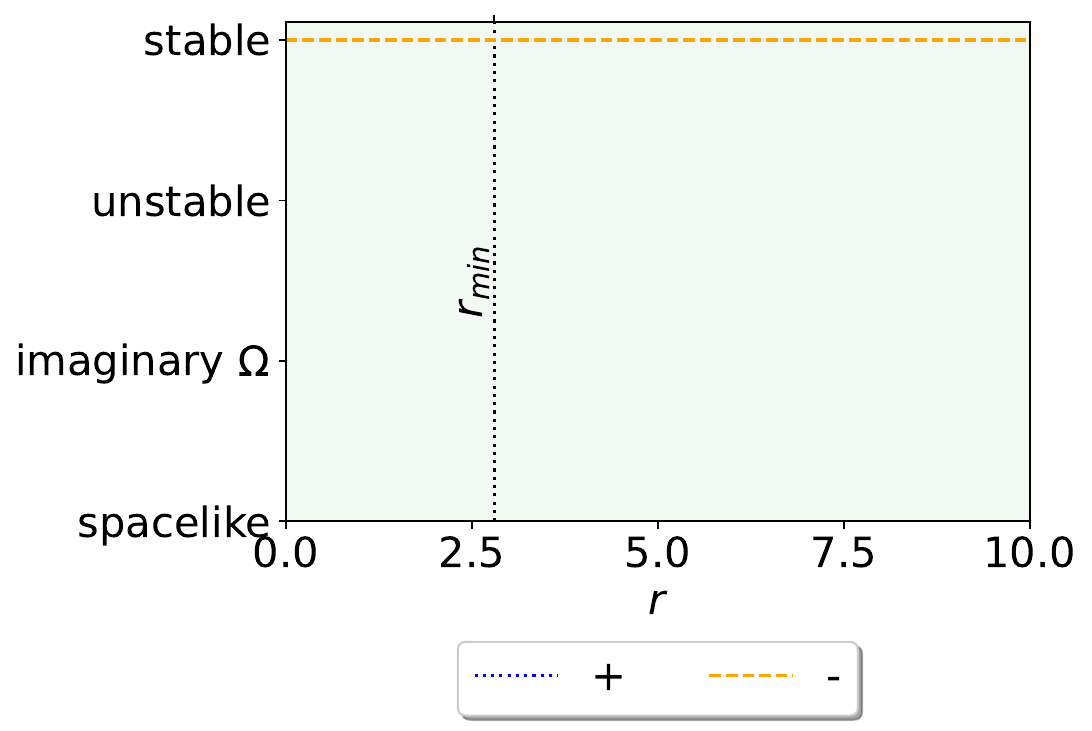}} 
		\subfloat[Case B]{\includegraphics[width=0.45\columnwidth]{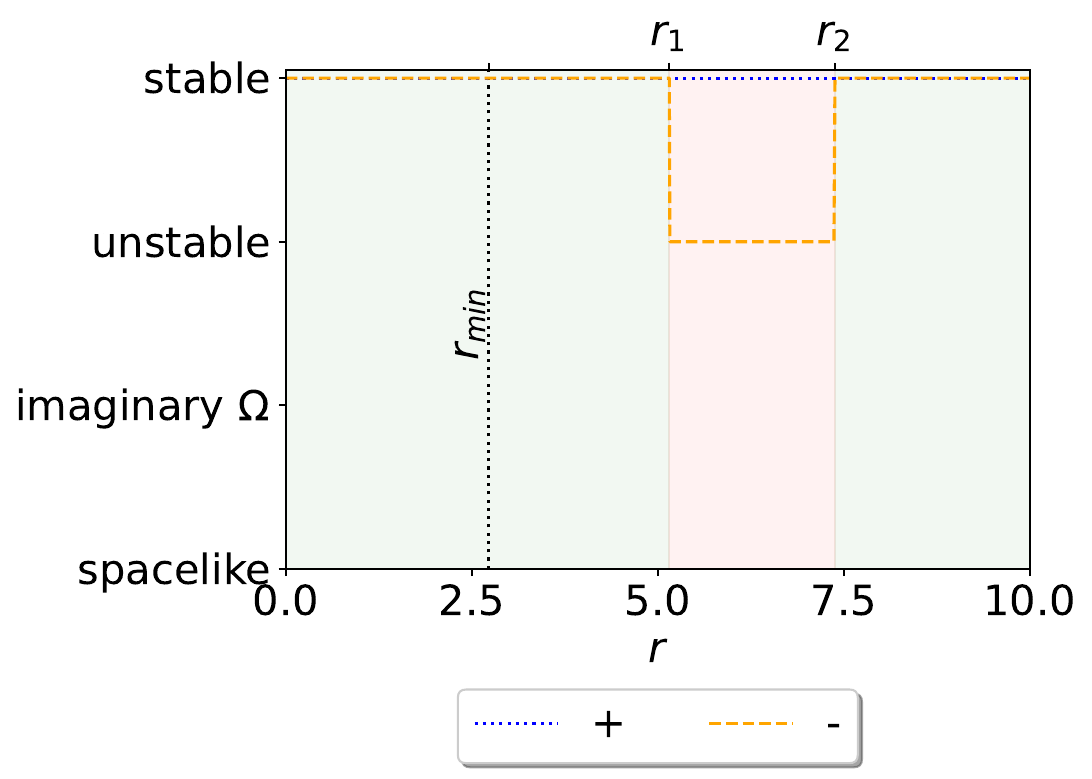}}\\
		\subfloat[Case C]{\includegraphics[width=0.45\columnwidth]{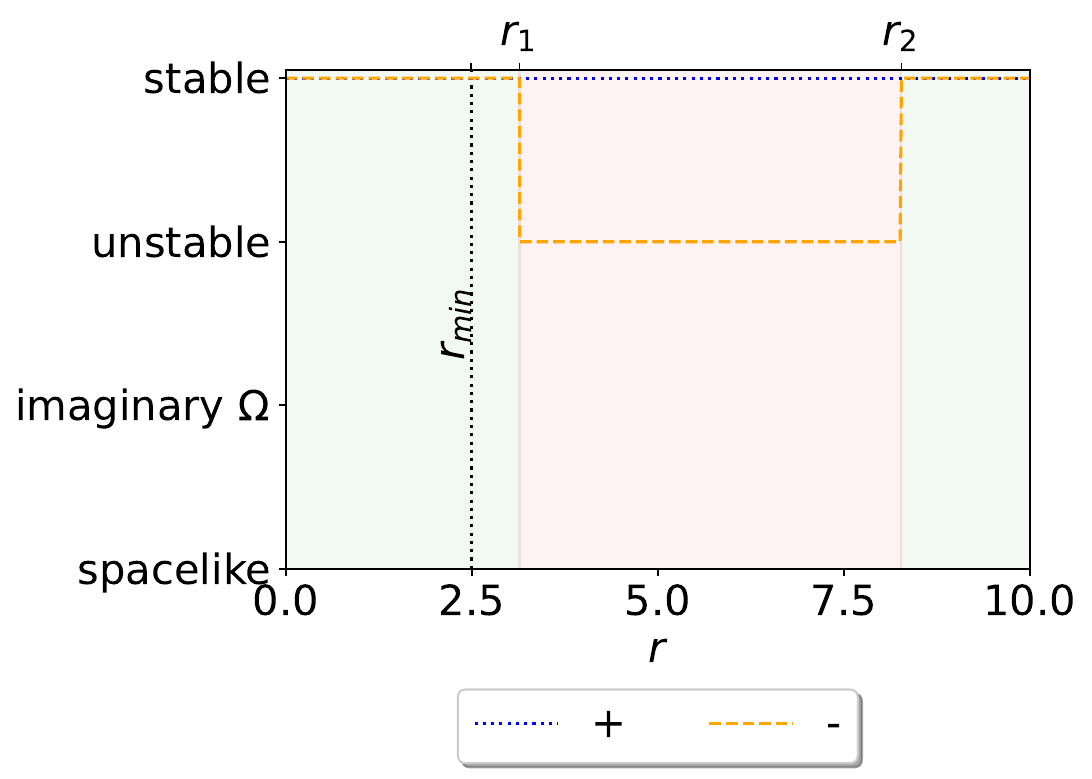}} 
		\subfloat[Case D]{\includegraphics[width=0.45\columnwidth]{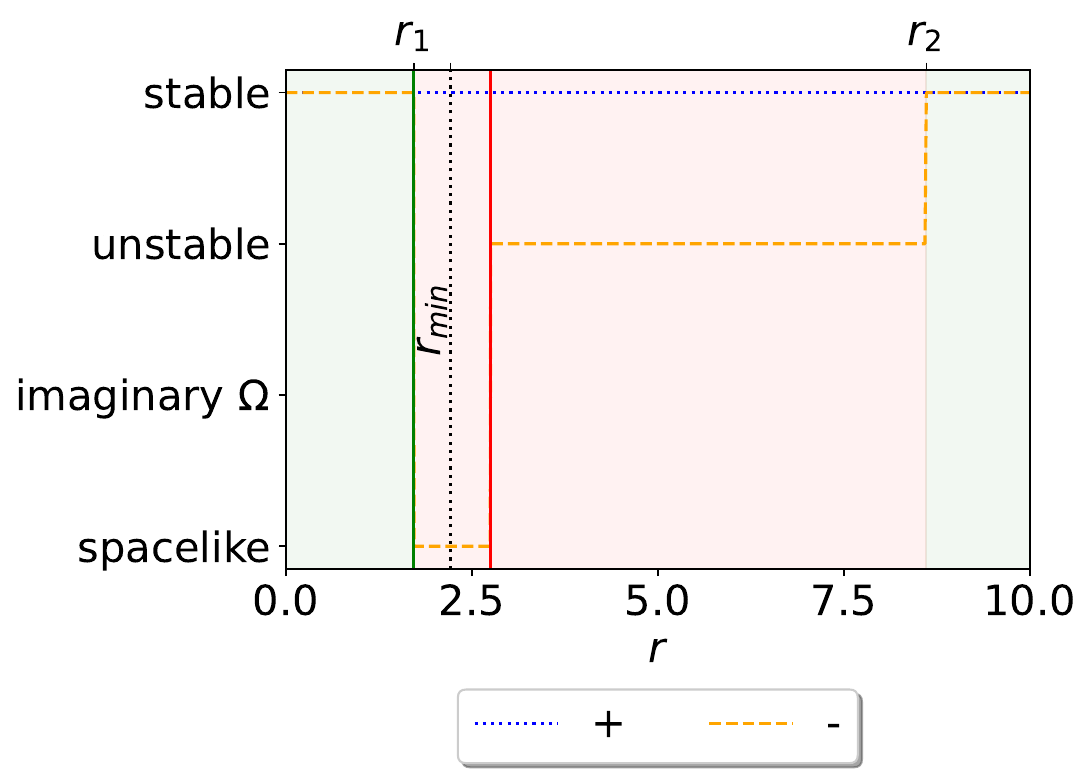}}
	\end{tabular}
	\caption{\small \label{fig:TCO-structure} TCOs structure for the four cases A, B, C and D, in Fig. \ref{fig:solution-space}. Horizontal orange lines correspond to the counter-rotating case, whilst horizontal blue lines correspond to the co-rotating one. For the counter-rotating case, stable TCO exist in the regions represented by a light green colour, and are not possible in the light red region. The dashed vertical lines mark, for the counter-rotating case, the maximum of the angular velocity. Stable (unstable) light rings, whenever present, are represented by a green (red) vertical line.   }
 \label{fig:TCO-structure}
\end{figure}

\section{Thin disk profiles and placement}
\label{sec:Disk}

By inspecting Fig.~\ref{fig:TCO-structure}, it becomes clear that in all four scenarios an actual ISCO never exists: regardless of the rotation sense being considered there is always a stable TCO in the neighbourhood of the origin. Nevertheless, it is possible to have regions with stable TCOs that are {\it disconnected}, as illustrated by the counter-rotating sector for the cases $B,C,D$. In sharp contrast, the co-rotating sector the stable TCOs always extend uninterrupted all the way up to infinity. For this reason, unless otherwise stated, we will focus mainly on the much richer counter-rotating sector of circular geodesics.\\

\begin{figure}[H]
	\begin{tabular}{cc}
		\subfloat[Case A]{\includegraphics[width=0.45\columnwidth]{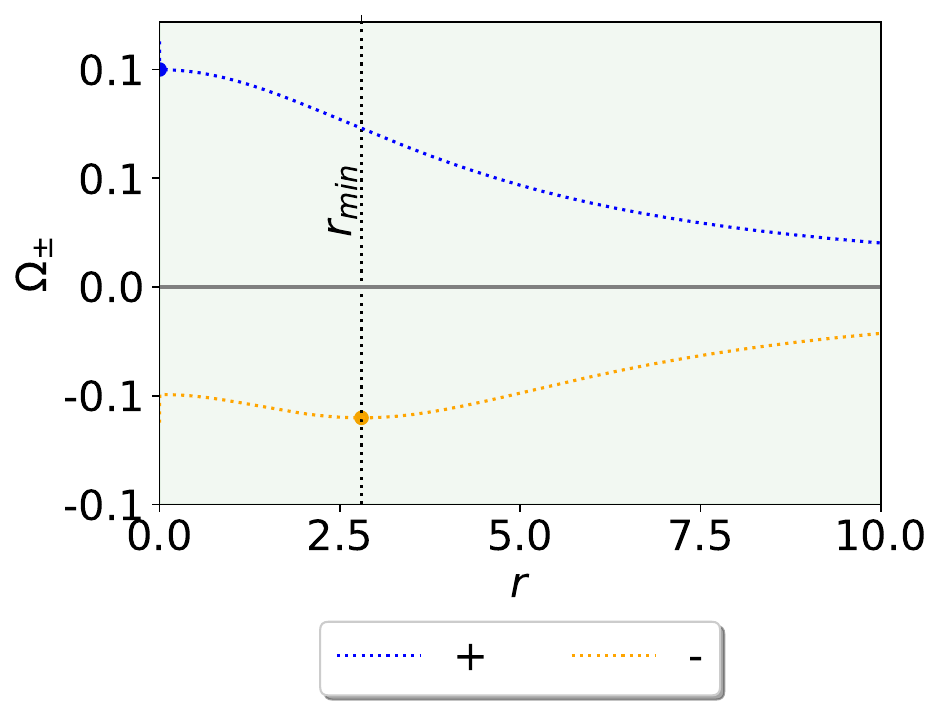}} 
		\subfloat[Case B]{\includegraphics[width=0.45\columnwidth]{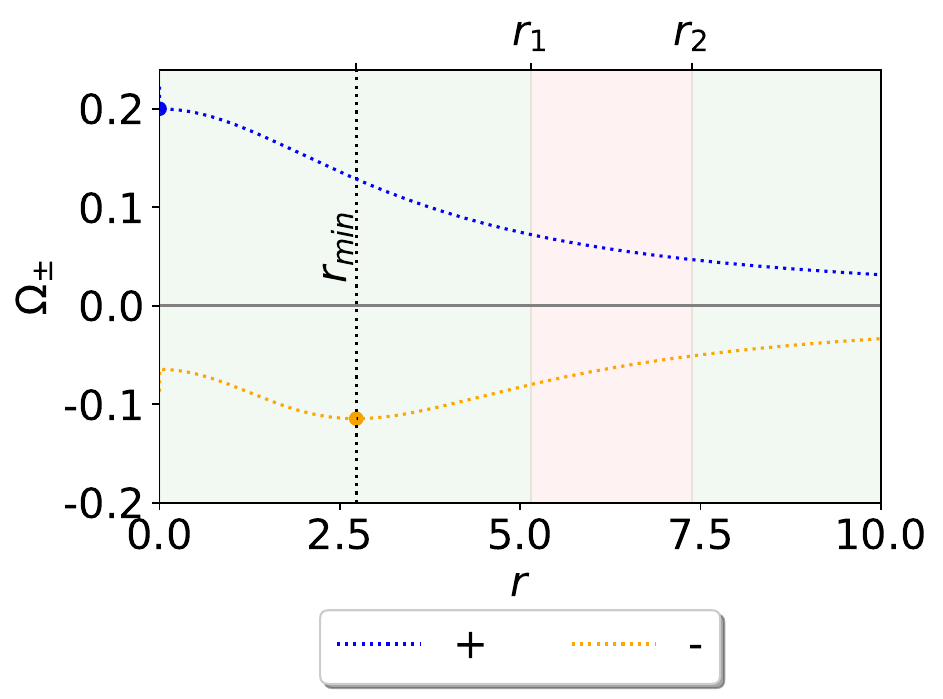}}\\
		\subfloat[Case C]{\includegraphics[width=0.45\columnwidth]{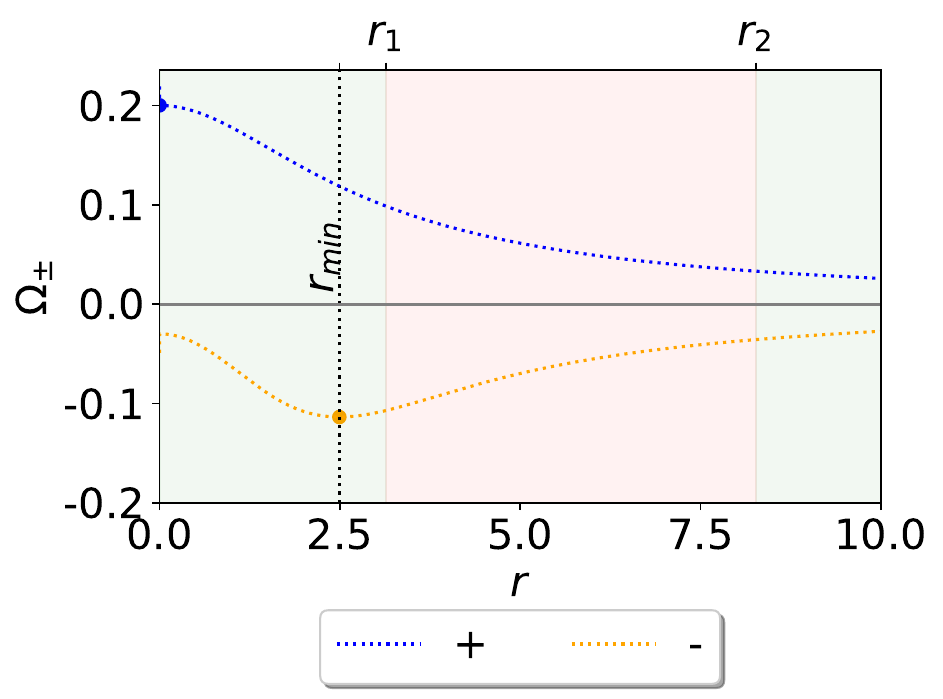}} 
		\subfloat[Case D]{\includegraphics[width=0.45\columnwidth]{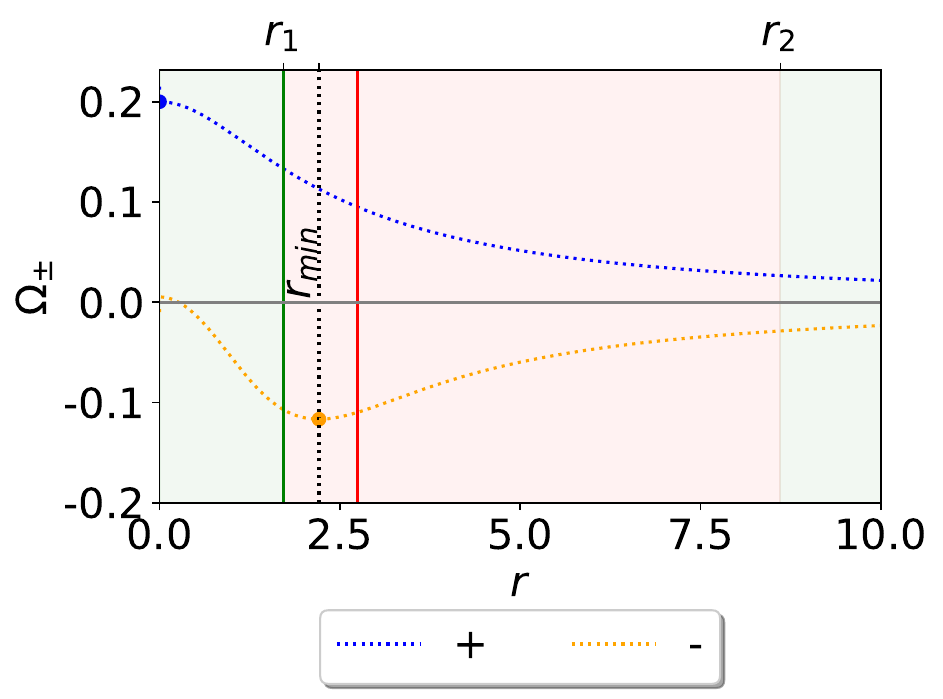}}
	\end{tabular}
	\caption{\small \label{fig:Omega-structure} Angular velocity  for the cases A, B, C and D, in Fig.\ref{fig:solution-space}. Orange curves correspond to the counter-rotating case, whilst the blue ones correspond to the co-rotating one. For the counter-rotating case, stable TCO exist in the regions represented by a light green colour, and are not possible in the light red region. The dashed vertical lines mark, for the counter-rotating case, the maximum of the angular velocity. Stable (unstable) light rings, whenever present, are represented by a green (red) vertical line. }
\end{figure}

Although the solutions $\{A,B,C,D\}$ do not feature an ISCO, similarly to the non-rotating PSs in~\cite{Herdeiro:2021lwl}, they might still mimic the appearance of a BH. This possibility arises from the existence of a maximum of the angular velocity $|\Omega|$ at a radius $r_{min}$, outside the PS's center (cf. Fig. \ref{fig:Omega-structure}). This notion draws inspiration from the GRMHD simulations reported in~\cite{Olivares:2018abq}, wherein the authors discuss how under certain assumptions, namely the loss of angular momentum in orbiting matter being mainly driven by the magneto-rotational instability, $r_{min}$ serves as a robust predictor for the inner edge of the accretion disk.

In light of the discussion above, and taking into account the intricate TCOs distributions presented in Fig.~\ref{fig:TCO-structure}, we can consider a number of possibilities regarding the placement of a disk. For concreteness, we shall consider geometrically thin disks on the equatorial plane, endowed with counter-rotating motion with respect to the central PS. One plausible formation mechanism for such a disk might be the tidal fragmentation of a main-sequence star that ventures too close  to the PS with negative orbital angular momentum. Such a process would lead to the creation of an accretion disk also with negative angular momentum, $i.e.$ counter-rotating with respect to the PS. The radial coordinate of the inner edge of the disk $R_0$ can be located at either $r_{min}$ or $r_2$, depending on the scenario considered. In addition, we will also assume one of the two luminosity profiles $I_0$ defined below:

\begin{align}
\label{eq_luminosity}
\begin{cases}
	&\text{``power decay" profile:} \quad I_0(r,R_o)=\frac{R_0/\mu^2}{r^2} \, ,  \\
&\text{``exponential decay" profile:} \quad I_0(r,R_o)=\mathrm{e}^{-10(r-R_0)\mu}.
\end{cases}
\end{align}

To conclude this section, we present for reference the values of the coordinates $r_{min}$, $r_1$ and $r_2$ (when applicable) in Table \ref{tab1}, in units of $\mu$, for the four case analysed.

 \begin{table}[H]
 	\centering
 	\begin{tabular}{l|c|c|c}
 		\hline
 		\hline
 		Case & $r_{min} \mu$ &$r_1 \mu$ & $r_2 \mu$   \\ \hline
 		A  & 2.8066             & N/A         & N/A     \\ \hline
 		B & 2.7223             &5.1496        & 7.4027      \\ \hline
 		C & 2.4948            & 3.1410        & 8.2746    \\ \hline
 		D & 2.2114             & 1.7194         & 8.6008      \\ \hline
 	\end{tabular}
 	\caption{\label{tab1} Coordinates $r_{min}$, $r_1$ and $r_2$ (when applicable), in units of $\mu$}
 \end{table}

\section{Lensing}
\label{sec:lensing}

\begin{figure}[H]
	\centering
	\begin{tabular}{cc}
		\subfloat[Case $A$]{\includegraphics[width=0.3\columnwidth]{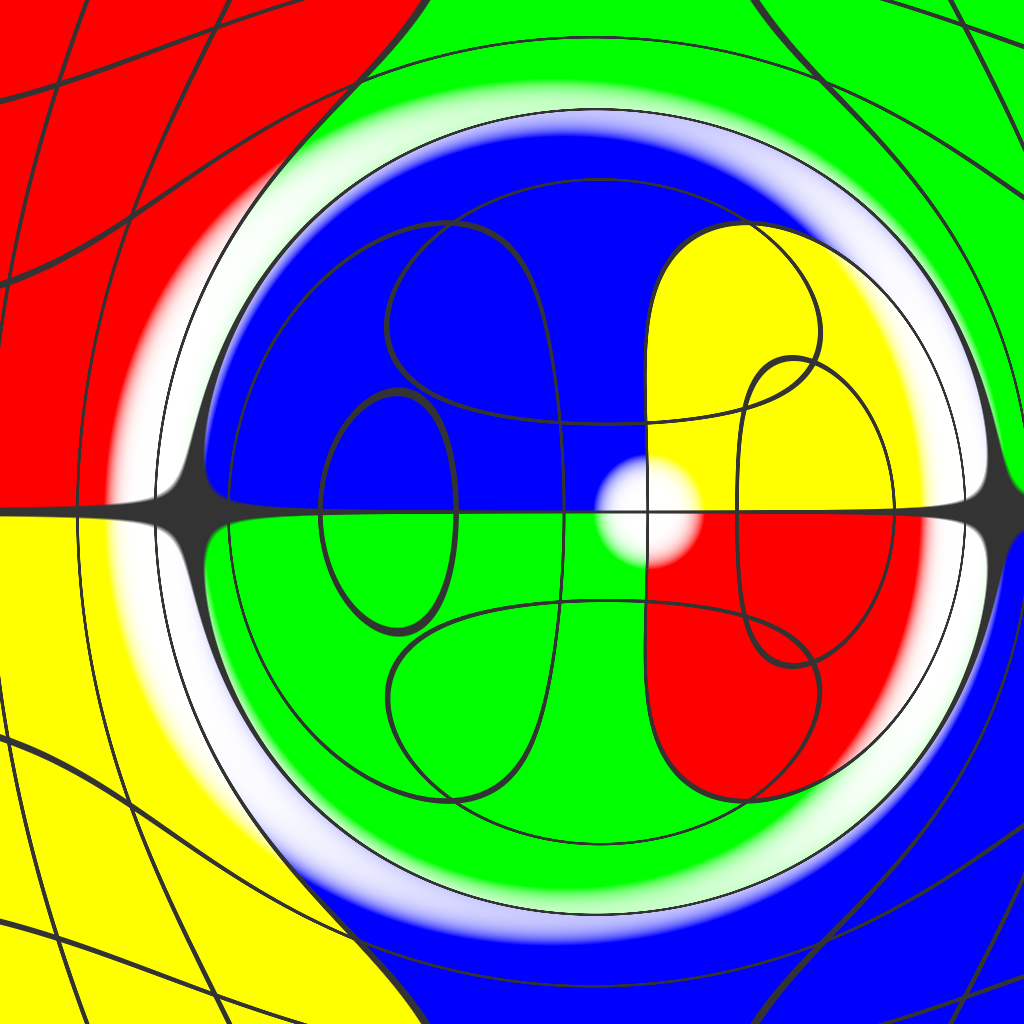}} 
		\subfloat[Case $B$]{\includegraphics[width=0.3\columnwidth]{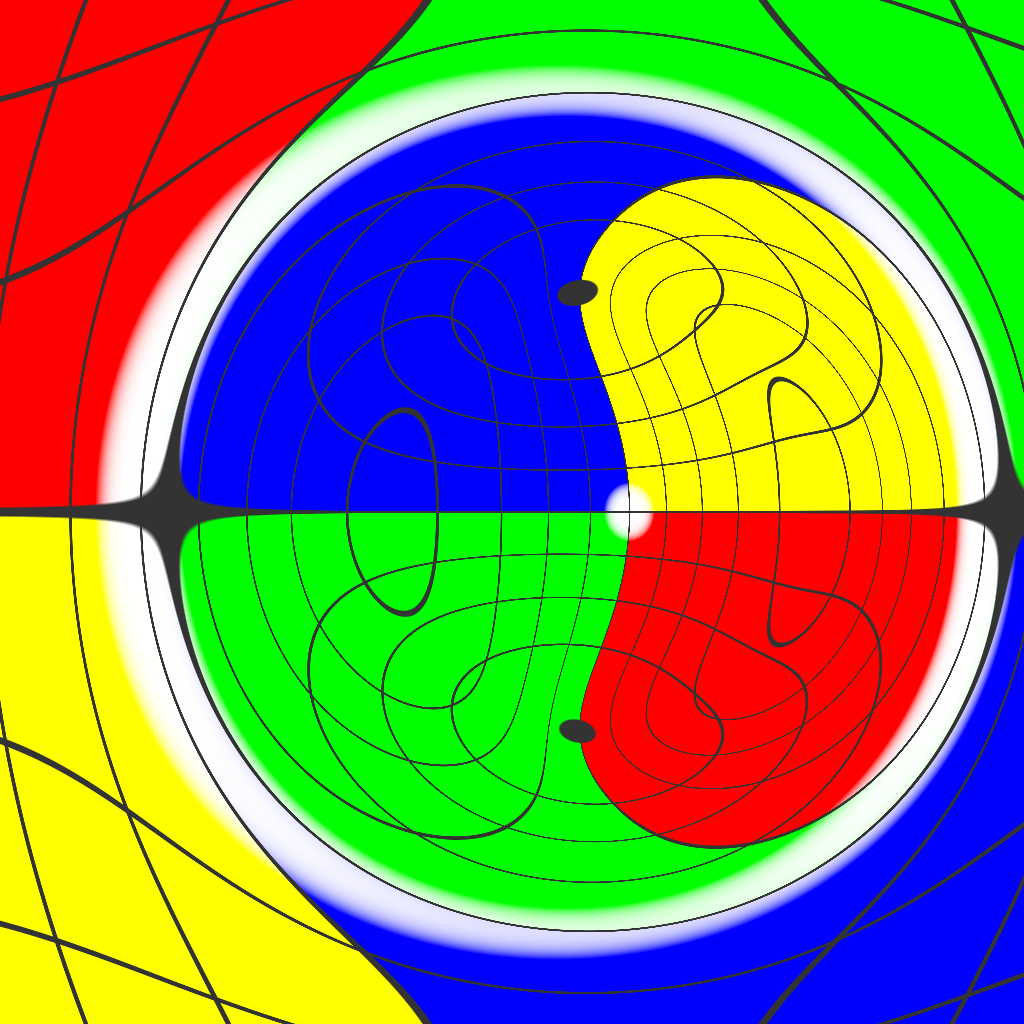}}\\
		\subfloat[Case $C$]{\includegraphics[width=0.3\columnwidth]{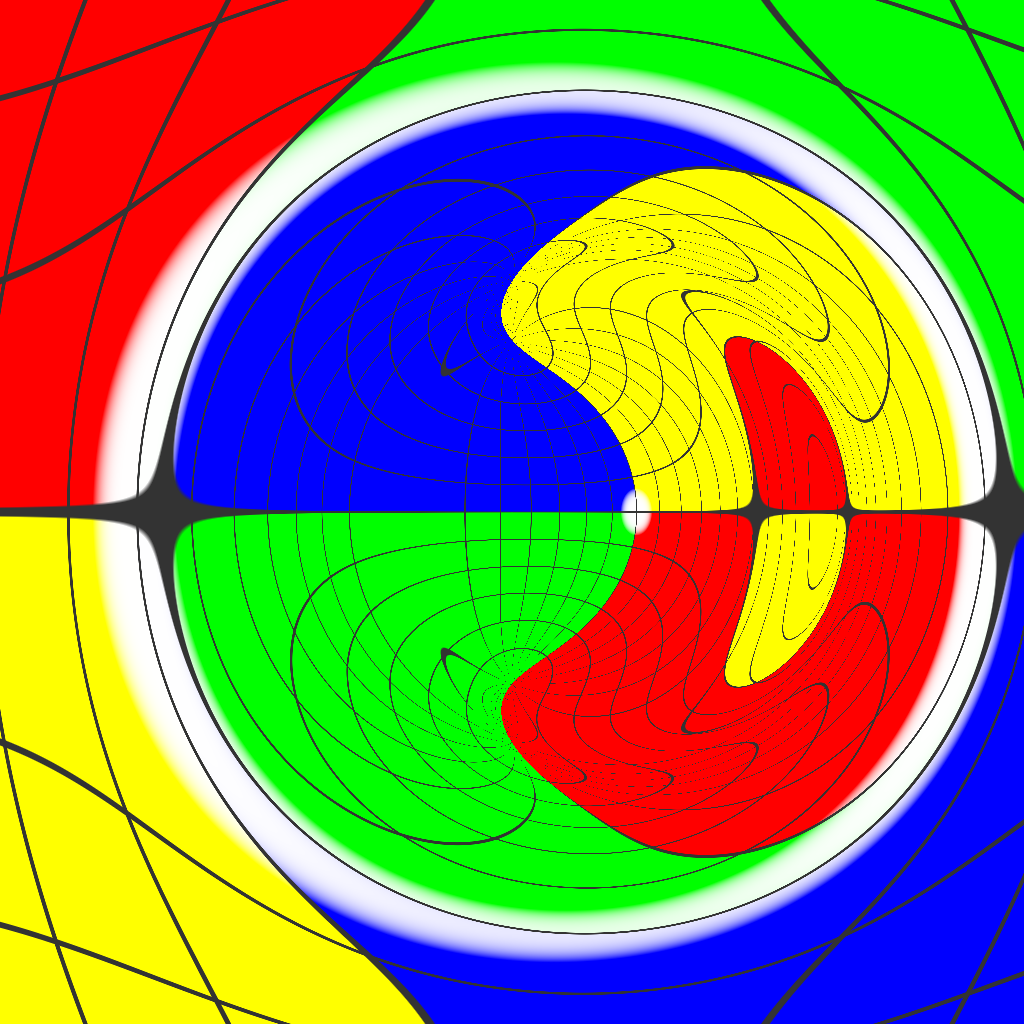}} 
		\subfloat[Case $D$]{\includegraphics[width=0.3\columnwidth]{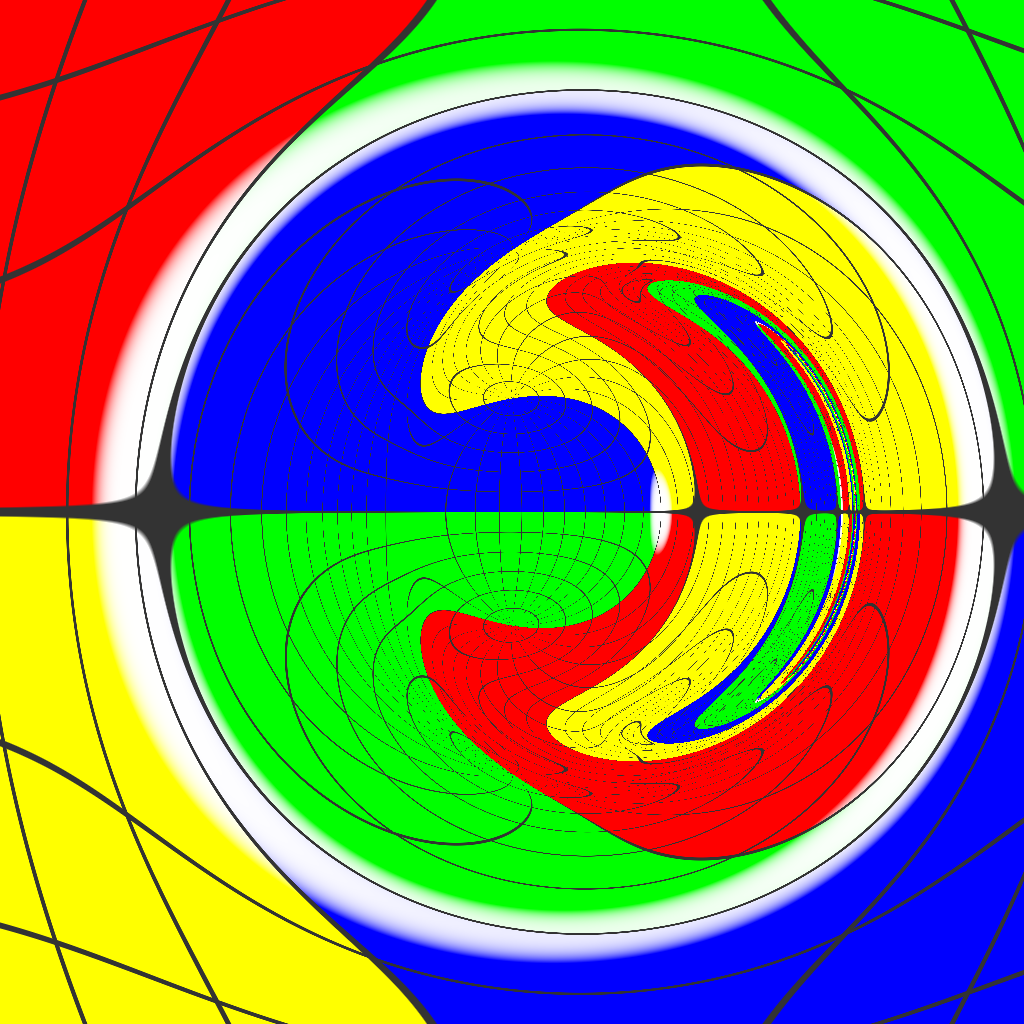}}
	\end{tabular}
	\caption{\small \label{fig:academic-lensing} Lensing images obtained with a bright celestial sphere. Adapted from~\cite{Sengo:2022jif}.  }
\end{figure}

Academic lensing images for solutions $A-D$ have already been thoroughly examined in~\cite{Sengo:2022jif},  wherein an academic bright celestial sphere was employed as the light source (see also~\cite{Bohn:2014xxa}). Although this setup is astropysically unrealistic, it serves as a valuable tool for understanding light-bending phenomena within the spacetime. For reference, in Fig.~\ref{fig:academic-lensing} we reproduce these more academic images, which will offer some insights into the intricate gravitational lensing effects associated with PSs.\\

In this study, we transition to a more astrophysical setup by adopting a luminous accretion disk as the light source surrounding the compact object. We employ a ray-tracing procedure akin to that discussed in~\cite{Cunha:2016bpi}, but now all pixels corresponding to trajectories that do not intersect the disk are rendered black. In contrast, pixels intersecting the disk at some point are represented in varying shades of red, defined by the normalized~\footnote{The RGB array $[Red,Green,Blue]$ is normalized with respect to the white color, for which all components are unity, $i.e.$ $[1,1,1]$. Black corresponds then to $[0,0,0]$.} RGB array $[I_0(r^*),0,0]$. Here, $I(r^*)$ denotes the intensity profile defined by Eq.~\ref{eq_luminosity}, evaluated at intersection point of the light ray with the disk, at the radial coordinate $r^*$.
Regarding the observation frame, and similarly to~\cite{Cunha:2016bpi}, we will be considering a zero axial angular momentum observer (ZAMO) with respect to spatial infinity (see Appendix~\ref{App1}). By default, the observer will be set at a circumferential radius  $r_{circ}=15M_{ADM}$.

\subsection{Case $A$}

\begin{figure}[H]
	\centering
	\begin{tabular}{cc}
		\subfloat[$\theta=5^o$ ]{\includegraphics[width=0.45\columnwidth]{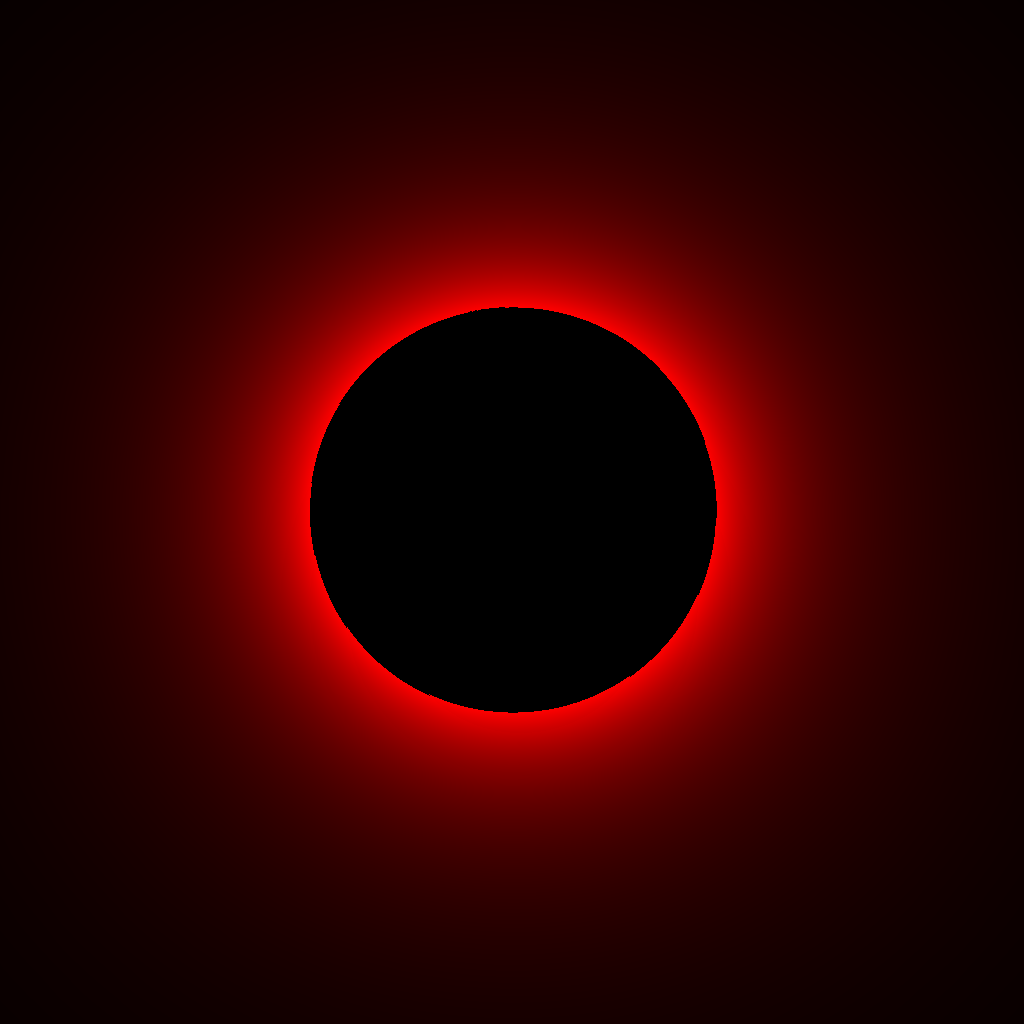}} 
		\subfloat[$\theta=88^o$ ]{\includegraphics[width=0.45\columnwidth]{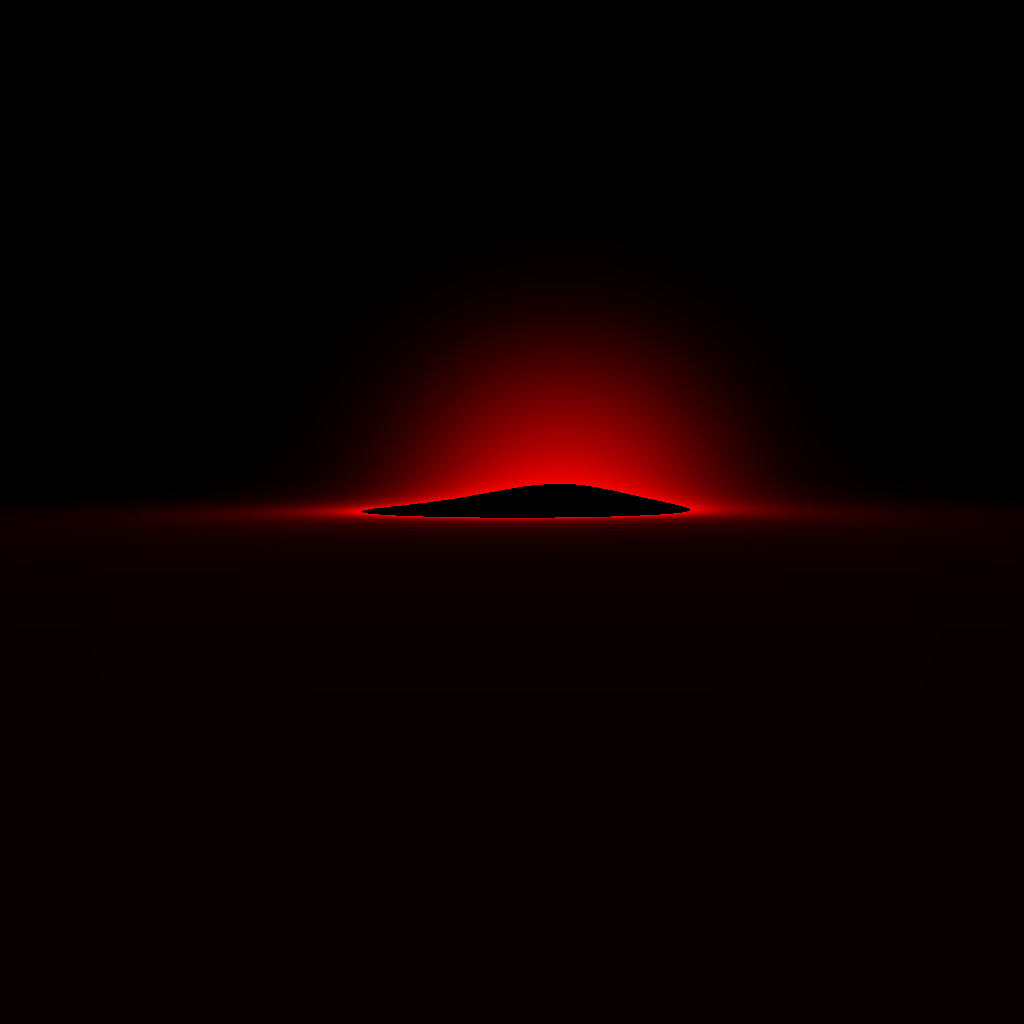}}
	\end{tabular}
	\caption{\small \label{fig:case-A} Lensing images for case A, for an observer at $\theta=5^o$ (left) and for $\theta=88^o$ (right). A ``power decay'' intensity profile was considered here (cf. Eq.~\ref{eq_luminosity}).}
\end{figure}

For the case of solution $A$, we position the inner edge of the disk at $r_{min}$. Given the stability of TCOs throughout the spacetime, the disk extends from $r_{min}$ to infinity, albeit with negligible luminosity beyond a certain distance. Displayed in Fig.~\ref{fig:case-A} are the lensing images for solution $A$, obtained with the ``power decay'' intensity profile (cf. Eq.~\ref{eq_luminosity}), under both face-on (left) and edge-on (right) observation angles. The face-on observation reveals a lensing pattern reminiscent of a Schwarzschild BH. However, similarly to the solutions considered in~\cite{Herdeiro:2021lwl}, this resemblance is less striking when viewed from an angle close to the equator. Indeed, due to the lower compactness of solution $A$, the lensing effect is not strong enough to replicate the characteristic profile associated with Schwarzschild BH lensing, as discussed for instance in~\cite{Herdeiro:2021lwl,Rosa:2022tfv}.

\subsection{Case $B$ }

\begin{figure}[H]
	\centering
	\begin{tabular}{cc}
		\subfloat[\scriptsize inner disk at $r_{min}$ (power decay) ]{\includegraphics[width=0.33\columnwidth]{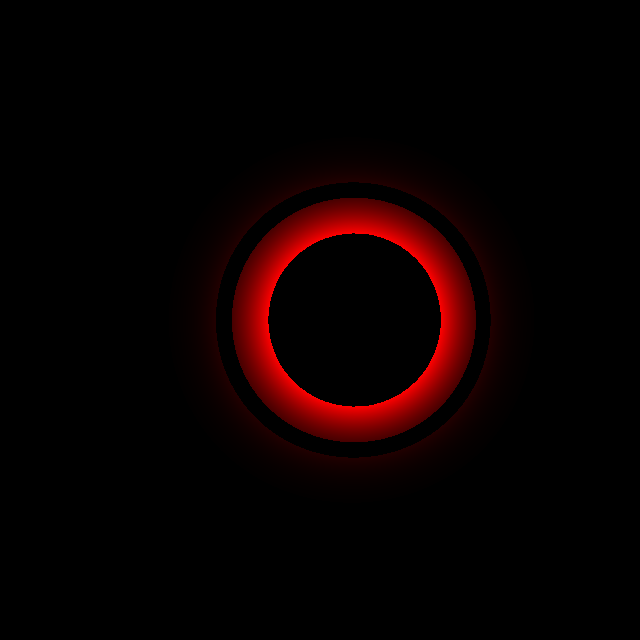}} 
		\subfloat[\scriptsize inner disk at $r_2$ (power decay) ]{\includegraphics[width=0.33\columnwidth]{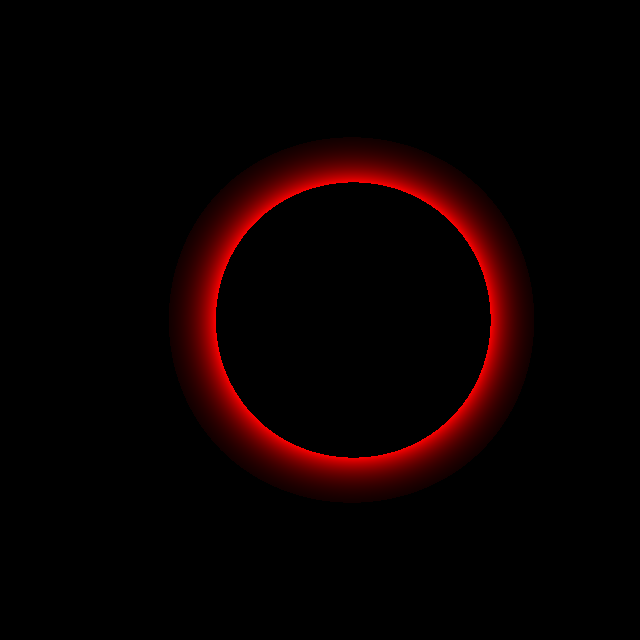}}\\
		\subfloat[\scriptsize double-disk with power decay profile]{\includegraphics[width=0.33\columnwidth]{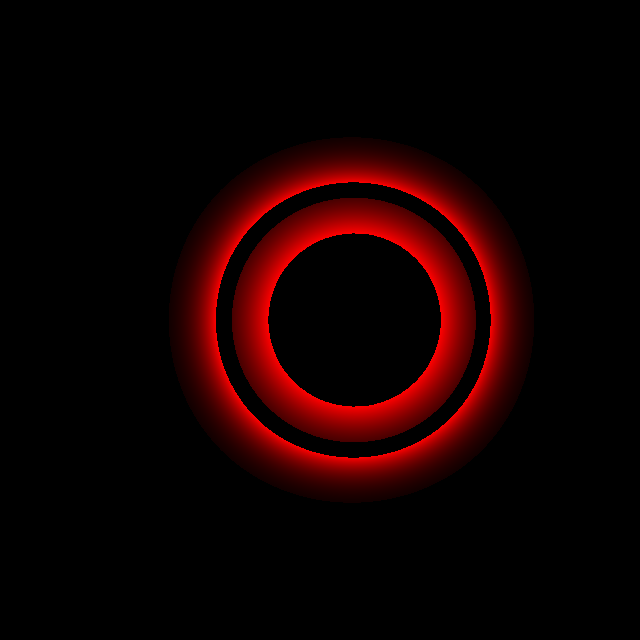}}
		\subfloat[\scriptsize double-disk with mixed decay profile]{\includegraphics[width=0.33\columnwidth]{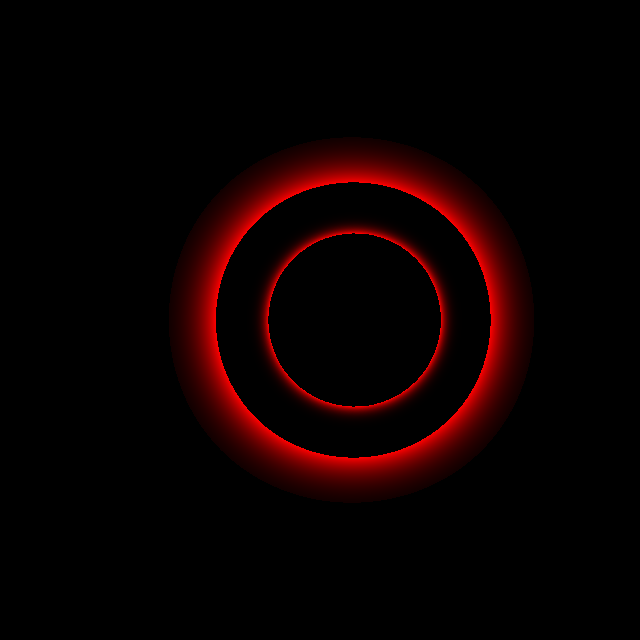}}
	\end{tabular}
	\caption{\small \label{fig:case-B} Lensing images for solution $B$, for an observer at $\theta=90^o$. }
\end{figure}

Continuing with case $B$, we initiate our exploration with a ``power decay'' intensity profile and by positioning the inner edge of the disk at $r_{min}$. We assume that the disk occupies solely those regions in the spacetime where stable TCOs are permitted. Consequently, any luminosity contributions from geodesics that exclusively intersect the equatorial plane between $r \in \left[r_1, r_2 \right]$ are disregarded, corresponding to black pixels in the resulting image. Illustrated in the top-left panel of Fig.~\ref{fig:case-B}, denoted as case (a), is the lensing image generated under this configuration when viewed from the equator. 

\begin{figure}[H]
	\centering
	\begin{tabular}{cc}
		\subfloat[$\theta=5^o$ ]{\includegraphics[width=0.3\columnwidth]{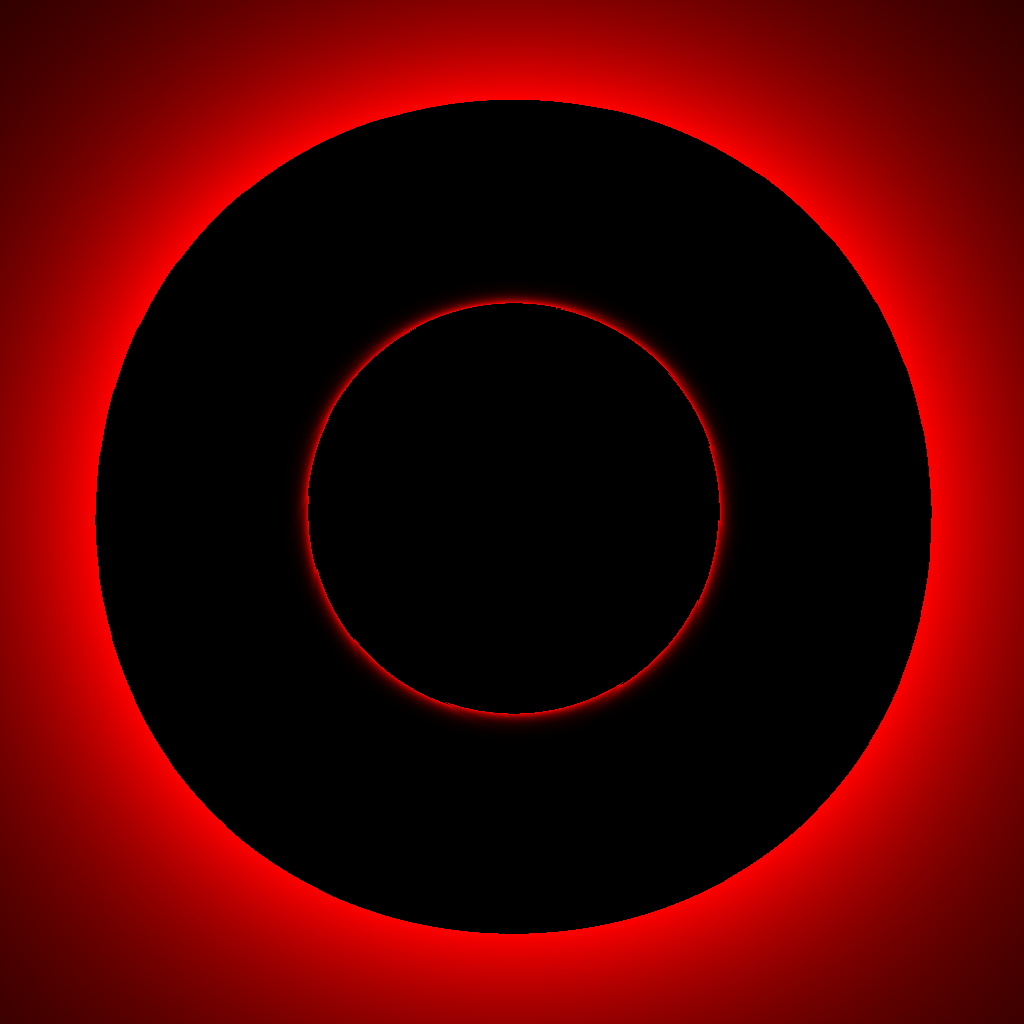}} 
		\subfloat[$\theta=88^o$ ]{\includegraphics[width=0.3\columnwidth]{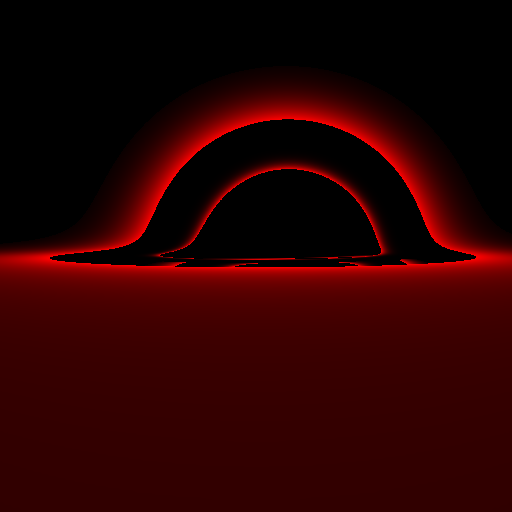}}\\
	\end{tabular}
	\caption{\small \label{fig:case-B2} Lensing images for case $B$ slightly off the equatorial plane (axis).}
\end{figure}

Nevertheless, this might not be immediately apparent. Indeed, upon comparing the images presented in Fig.~\ref{fig:case-A} with those in Fig.~\ref{fig:case-B}, the latter images might mistakenly appear to have been obtained in the proximity of the rotation axis, rather than at the equatorial plane. This misconception arises as a consequence of the infinitesimally thin approximation used in modelling the disk, together with the observer being placed {\it exactly} at the equatorial plane. Consequently, the direct image of the disk line is also infinitesimally thin. To clarify this, in one of the images of Fig.~\ref{fig:case-B2}, we deliberately shift the observer slightly away from the equatorial plane. \\

What if we change the location of the inner edge of the disk?  The top-right panel of Fig.~\ref{fig:case-B}, case (b), displays the lensing image corresponding to an accretion disk with its inner edge positioned at $r_2$, employing also a ``power decay'' intensity profile. The only discernible difference from the previous scenario lies in the size of the effective shadow, which expands noticeably with the relocation of the inner disk radius from $r_{min}$ to $r_2$.

Since in solution $B$ there are two disconnected regions with stable TCOs, we can consider a scenario where both regions have disks with independent emission profiles, $i.e.$ a double-disk. The latter is illustrated on the left bottom panel of Fig. \ref{fig:case-B}, case (c), with the emission given by:

\begin{align}
\label{eq_luminosity}
\begin{cases}
& \text{a `` power decay'' profile $I_o(r,r_{min})$ for $r\in\left[r_{min}, r_1 \right]$} ,  \\
&\text{a `` power decay'' profile $I_o(r,r_2)$ for $r\in\left[r_2, +\infty \right[$}.
\end{cases}
\end{align}

What if the luminosity decay rate is mixed between the disconnected sections of the double-disk? On the right bottom panel \ref{fig:case-B}, case (d), we explore such a case: we consider a (faster) exponential decay profile for the inner region of the disk, and a (slower) power decay profile for the outer disk region:

\begin{align}
\label{eq_luminosity2}
\begin{cases}
& \text{an ``exponential decay'' profile $I_o(r,r_{min})$ for $r\in\left[r_{min}, r_1 \right]$},  \\
&\text{a ``power decay'' profile $I_o(r,r_2)$ for $r\in\left[r_2, +\infty \right[$} .
\end{cases}
\end{align}

Such a mixed profile leads to the creation of a sharp feature in the image, which is typically found around BHs. It has been established in the literature~\cite{Olivares:2018abq,Herdeiro:2021lwl,Rosa:2022tfv} that a bosonic star, when accompanied by an accretion disk truncated at a specific radial coordinate away from the origin, can overall replicate the shadow of a BH. However, as discussed in~\cite{Herdeiro:2021lwl}, these objects' lensing images typically lack the prominent bright emission ring characteristic of their BH counterparts, which stems from the existence of a LR. We aim to highlight that in scenarios where stable but disconnected TCO regions exist, particularly if the innermost region is modeled by a sharp decay emission profile, then the resulting lensing image can exhibit a structure closely resembling the bright emission ring observed in the BH scenario. The resemblance to a BH image is even more striking in Fig. \ref{fig:case-B2}, where we present lensing images for observation angles of $5^o$ and $88^o$.

\subsection{Case $C$}

Both solutions $B$ and $C$ exhibit a region where only unstable (counter-rotating) TCOs exist, with the distinction being that this region is much larger in the case of $C$. However, given the more compact nature of solution $C$, it introduces further intriguing features that arise due to the intricate motion of light rays.\\

\begin{figure}[H]
	\centering
	\begin{tabular}{ccc}
		\subfloat[\scriptsize inner disk at $r_{min}$ (power decay) ]{\includegraphics[width=0.33\columnwidth]{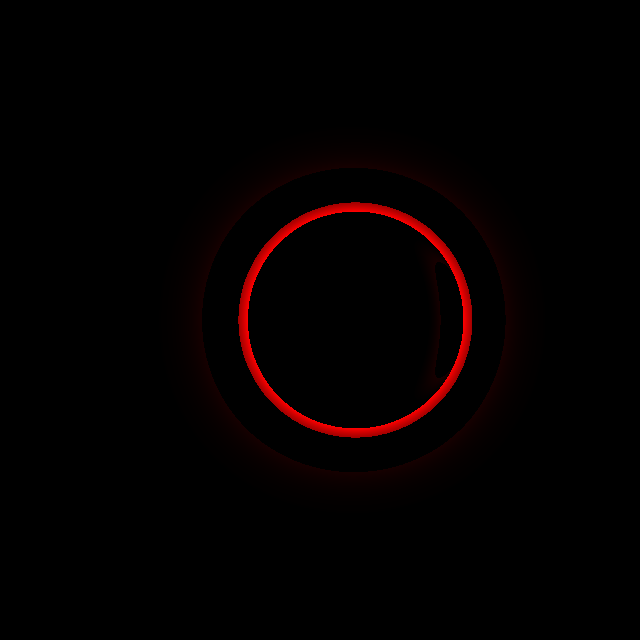}} 
		\subfloat[\scriptsize inner disk at $r_2$ (power decay)]{\includegraphics[width=0.33\columnwidth]{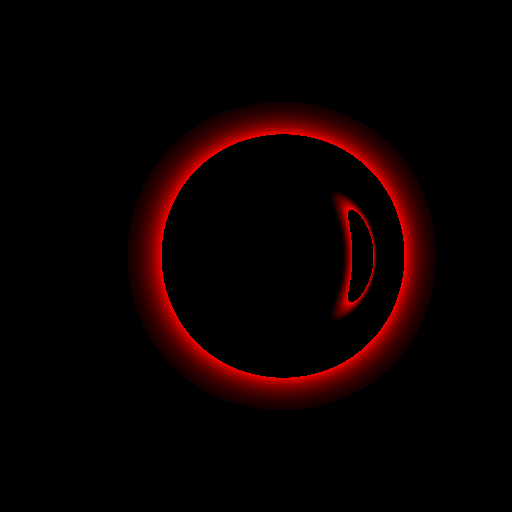}}
		\subfloat[\scriptsize double-disk (mixed decay)]{\includegraphics[width=0.33\columnwidth]{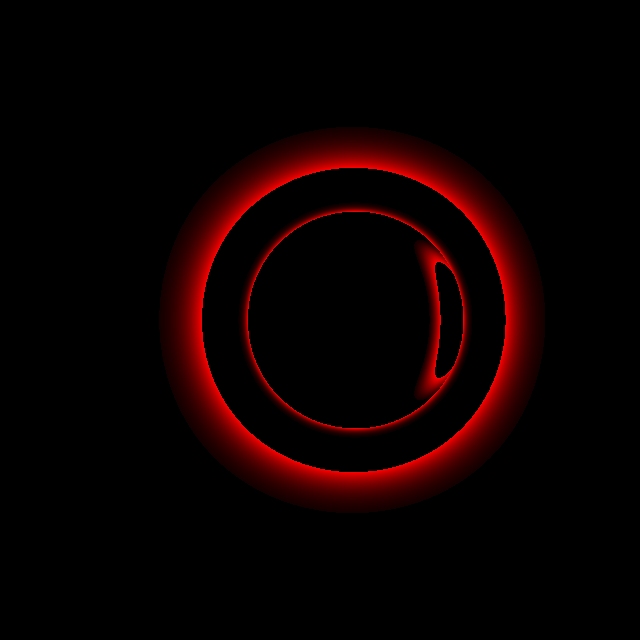}}
	\end{tabular}
	\caption{\small \label{fig:case-C} Lensing images for case $C$, for an observer at $\theta=90^o$.}
\end{figure}

In Fig.~\ref{fig:case-C}, mirroring our approach to case $B$, we present the observation image for an observer located at the equator, considering different disk placement combinations. Notably, in the middle panel of Fig.~\ref{fig:case-C}, we showcase the lensing image corresponding to an accretion disk with its inner edge positioned at $r_2$, featuring a ``power law'' luminosity decay. Despite the absence of emissions from regions below $r_2$, a luminous feature emerges within the effective shadow of the star. This feature corresponds to a double copy of certain sections of the disk. One can have a more intuitive understanding of this feature by comparison with the corresponding colored academic lensing image in Fig.~\ref{fig:academic-lensing}.

\begin{figure}[H]
	\centering
	\begin{tabular}{cc}
		\subfloat[$\theta=5^o$ ]{\includegraphics[width=0.3\columnwidth]{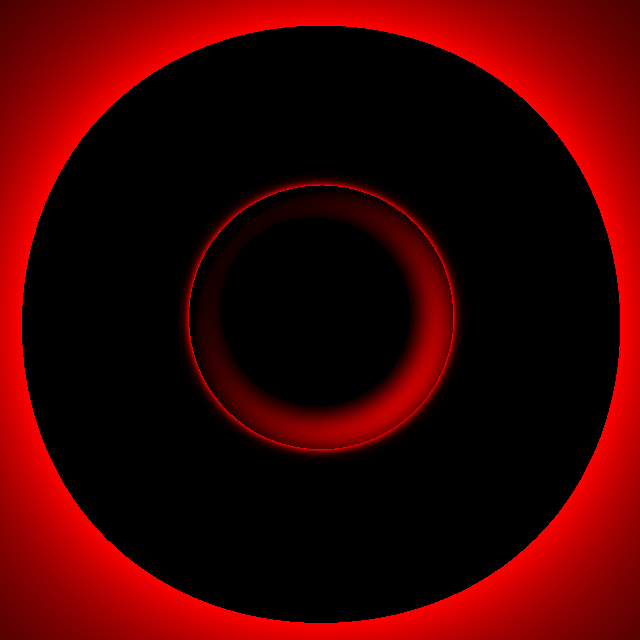}} 
		\subfloat[$\theta=88^o$ ]{\includegraphics[width=0.3\columnwidth]{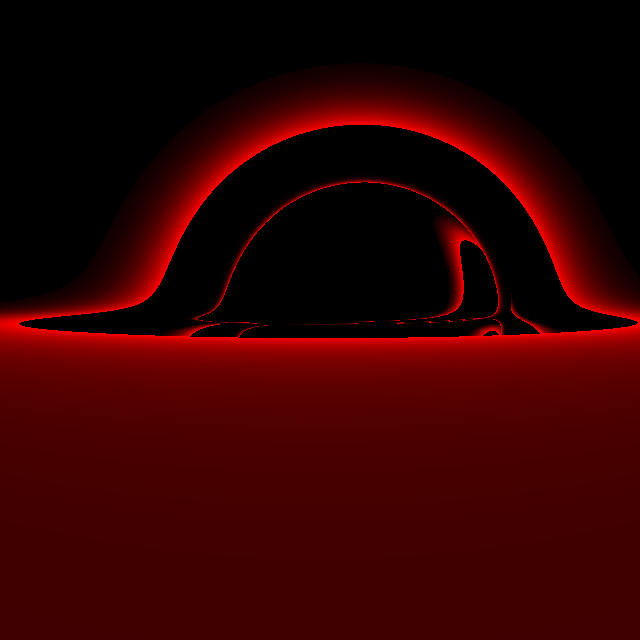}}\\
	\end{tabular}
	\caption{\small \label{fig:case-C2} Lensing images for case $C$ viewed slightly off the equatorial plane (axis).}
\end{figure}

Finally in Fig.~\ref{fig:case-C2} we present the double-disk scenario as viewed from $5^o$ and for $88^o$. 

\subsection{Case $D$}

Case $D$ is representative of solutions containing a pair of LRs -- one stable and one unstable. While the astrophysical relevance of this solution remains uncertain, as the presence of a stable LR is suspected to induce a spacetime instability~\cite{Cunha:2022gde,Keir:2014oka,Benomio:2018ivy}, it might be instructive to explore the potential impact of these LRs on the resulting lensing images.\\

\begin{figure}[H]
	\centering
	\begin{tabular}{cc}
		\subfloat[$\theta=5^o$ ]{\includegraphics[width=0.3\columnwidth]{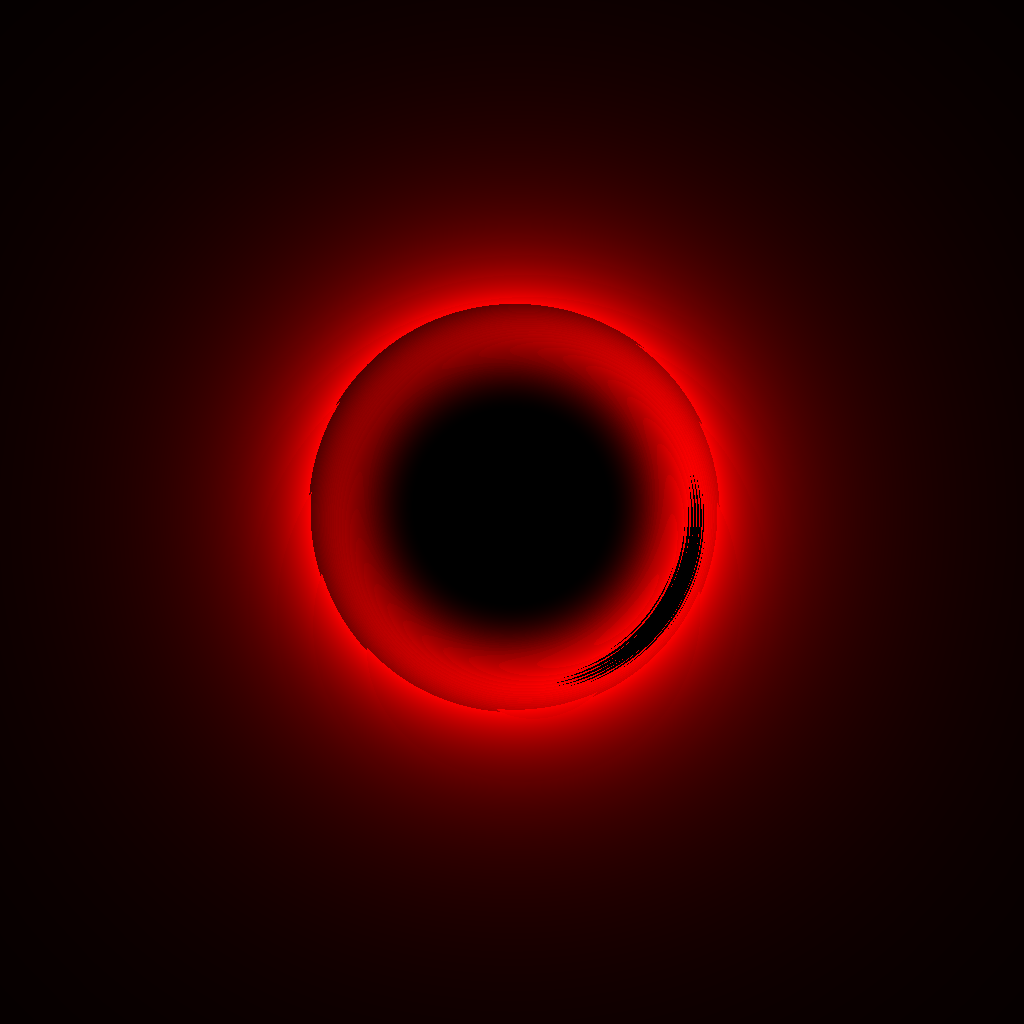}} 
		\subfloat[$\theta=88^o$ ]{\includegraphics[width=0.3\columnwidth]{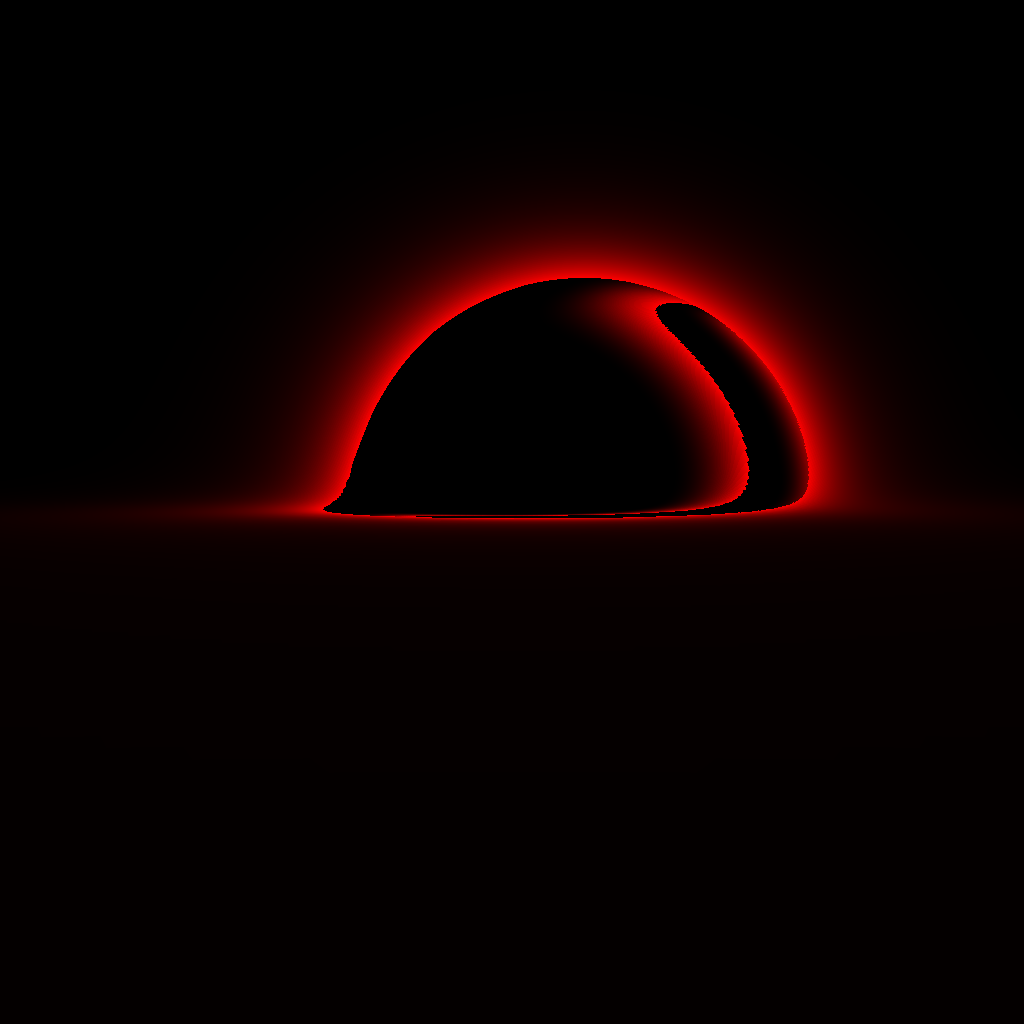}}
	\end{tabular}
	\caption{\small \label{fig:case-D} Lensing images for case $D$, viewed slightly off the equatorial plane (axis).}
\end{figure}

As detailed in~\cite{Delgado:2021jxd}, the region between LRs is devoid of any timelike orbits. The stable (unstable) LR's position is visually indicated in Fig. \ref{fig:TCO-structure} by a green (red) vertical line. Additionally, from Fig.~\ref{fig:Omega-structure}, it is clear that the extrema of the angular velocity occurs between the two LRs. Consequently, within the interval $(r_1, r_{min})$ it is not viable to place a stable accretion disk, as done previously. In Fig.~\ref{fig:case-D}, we showcase the lensing image for solution $D$, where we consider a ``power decay'' emission profile, with the inner edge positioned at $r_2$.

\section{Including redshift}
\label{sec:redshift}

A notable simplification in the images presented in the preceding section was the absence of gravitational redshift and Doppler effects, with the aim of emphasizing lensing features solely related to geodesic motion. 

In a more realistic scenario, consider a set of photons emitted from the disk with a frequency $\nu_0$ at some local co-moving disk frame $\mathcal{D}$. If these photons reach a distant observer $\mathcal{O}$, the observed frequency $\nu$ will generically be different than $\nu_0$. Moreover, and in addition to this frequency shift, the emission intensity $I_{\nu_0}$ of the radiation emitted locally at $\mathcal{D}$ will also differ from the detected intensity $I_\nu$ at $\mathcal{O}$. These transformations are intricately connected through the volume conservation in phase space for the considered set of photons, as described by Liouville's theorem~\cite{Misner2017}:

\begin{align}
	\label{eq_red}
	I_{\nu_0}= \left(\nu_0 / \nu  \right)^3 I_{\nu} \, .
\end{align}

For a ZAMO observer $\mathcal{O}$, one can show (see Appendix \ref{App1}) that 

\begin{align}\label{eq_red2}
	\frac{\nu_0}{\nu}= \left(\frac{1}{\zeta - \gamma \eta} \right)\Bigg|_{\mathcal{O}}  \left(1-\Omega \eta\right) \left(-g_{tt} -2 \Omega g_{t \phi} -\Omega^2 g_{\phi \phi}  \right)^{-1/2} \, .
\end{align}

In the latter expression, $\Omega$ is the disk's angular velocity at the location of $\mathcal{D}$, the photon's impact parameter $\eta=L_\infty/E_\infty$ is a geodesically conserved quantity defined by ratio between the photon's angular momentum $L_\infty$ and energy $E_\infty$ at infinity, while $\gamma$ and $\zeta$ depend on metric functions evaluated at $\mathcal{O}$:

\begin{equation}
    \gamma = -\frac{g_{t\phi}}{g_{\phi \phi}} \sqrt{\frac{g_{\phi \phi}}{g_{t \phi}^2 -g_{tt} g_{\phi\phi}}} \, , \quad \zeta=\sqrt{\frac{g_{\phi \phi}}{g_{t \phi}^2 -g_{tt} g_{\phi \phi}}}\,.\\
\end{equation}
The remaining metric functions in~\eqref{eq_red2} are evaluated at the disk point $\mathcal{D}$.\\

In order to obtain a lensing image in practice, multiple null geodesics are ray-traced backwards in time, starting from the observer $\mathcal{O}$'s location. Each light ray is characterised by a conserved impact parameter $\eta$, assigned based on the corresponding pixel in the observation image, and has a fixed frequency $\nu$ in the observation frame. Upon intersecting the accretion disk, a light ray establishes a point $\mathcal{D}$ as defined earlier. The projection of the photon's momentum onto the frame $\mathcal{D}$ allows the computation of the locally measured frequency $\nu_0$. Assuming that $I_{\nu_0}$ is a known property of the disk model, applying Eq.~\eqref{eq_red2} yields the intensity map $I_\nu$ for the initial image pixel.\\

\begin{figure}[H]
	\centering
	\begin{tabular}{cc}
		\subfloat{\includegraphics[width=0.30\columnwidth]{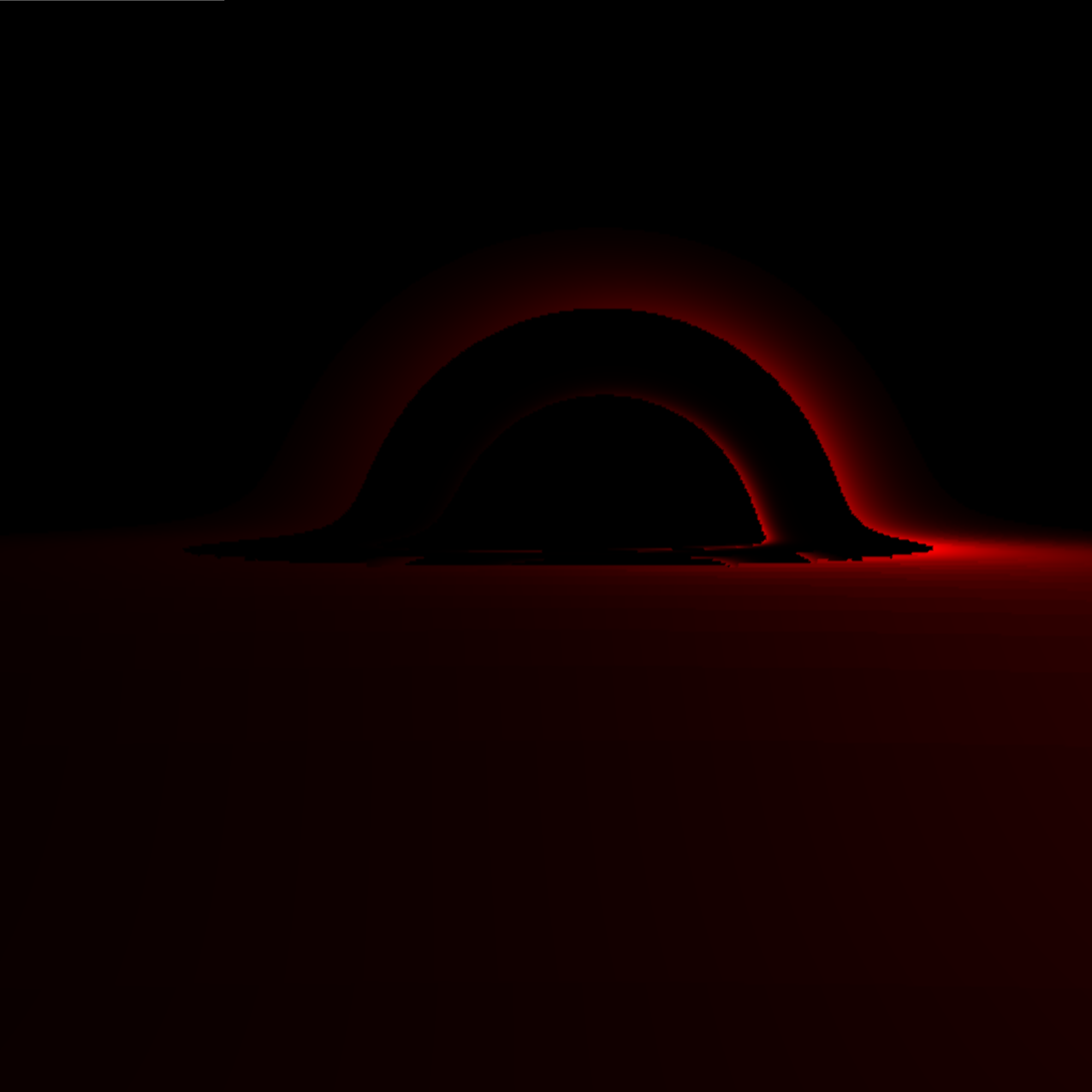}\includegraphics[width=0.30\columnwidth]{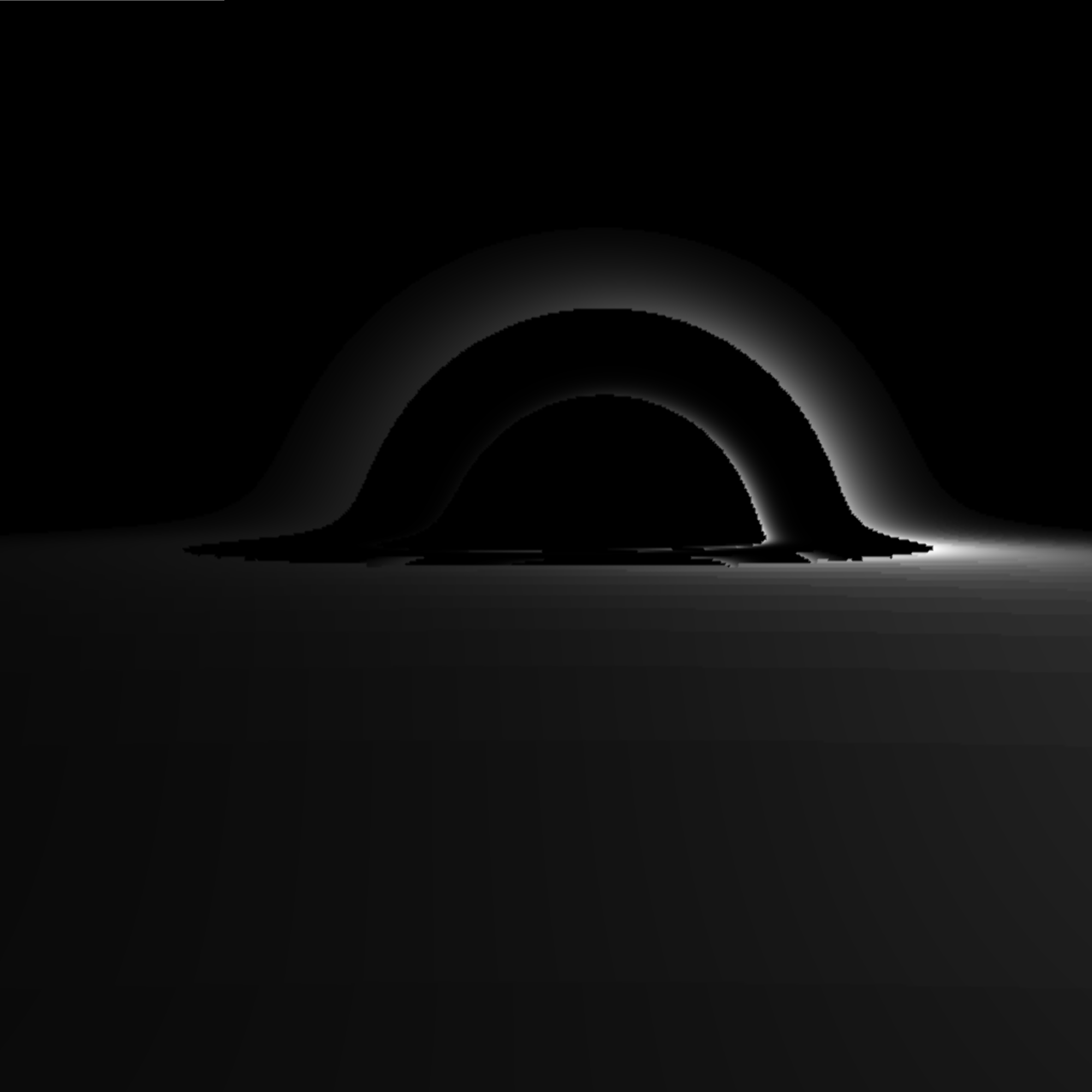}}
	\end{tabular}
	\caption{\small \label{fig:red-shift} Lensing image for model $B$ including a double disk model with a mixed intensity decay, now with the effects of redshift taken into account. The strong left-right image asymmetry is mainly due to the rotation of the disk. The image on the right is a black-and-white reproduction of the one on the left, given that the latter was originally quite dim.}
\end{figure}

In Fig.~\ref{fig:red-shift},  we revisit the lensing image of solution $B$ with a local luminosity decay rate $I_{\nu_0}$ provided by Eq.~\eqref{eq_luminosity2}, but now incorporating redshift effects. As before, this model is able to mimic the lensing image of a BH.\\

The intensity profile given by Eq.~\ref{eq_luminosity} is essentially ad-hoc, and was chosen precisely to highlight the BH mimicking properties of Proca stars. While similar outcomes could theoretically be achieved by employing a carefully selected temperature profile for a black body emission, the crucial question arises: Can such a temperature profile be realistically obtained under reasonable astrophysical conditions? The answer to this question might be obtained via GRMHD simulations, an analysis that we will consider in future works.

\section{Conclusions}
\label{sec:conclusions}

We present compelling evidence supporting the claim that PSs can mimic the visual appearance of BHs, when lit by accretion flow environments effectively described by a geometrically thin and optically thick accretion disk. Our findings address limitations identified in earlier studies of the Proca model. In particular, in contrast to the cases analysed in~\cite{Herdeiro:2021lwl}, we consider solutions that are more astrophysically viable, since they exhibit both rotation and dynamical stability. These solutions proved to be more capable of replicating characteristic features of BH images, due to three main ingredients:

\begin{enumerate}
	\item Similarly to~\cite{Herdeiro:2021lwl}, an effective shadow is achieved by introducing a cutoff in the emission profile of the accretion disk at the location of the maximum of the angular velocity (as explored in~\cite{Olivares:2018abq}). However, and in contrast to~\cite{Herdeiro:2021lwl}, the  presence of such a maximum outside the star center does not require the existence of self-interactions in the bosonic theory.
	\item The solutions considered here are more compact. Consequently, in contrast to~\cite{Herdeiro:2021lwl}, one can obtain lensing images of Proca stars that closely resemble those of BHs, even under edge-on observation scenarios.
	\item The complex structure of timelike geodesics in these spacetimes allows for various configurations for the placement of the disk. In principle, one could consider a scenario featuring a disk with two disconnected components. If the inner disk exhibits rapid emission decay compared to the outer one, this arrangement enables us to obtain features that closely resemble the thin bright emission ring that is typically associated to BH images.
\end{enumerate}  

The results presented in this study offer a preliminary proof of concept. To attain a deeper understanding, a more thorough analysis is essential, especially by considering the dynamics of more realistic accretion flows. Employing GRMHD simulations will be critical in confirming the applicability in the vector case of the observations made in~\cite{Olivares:2018abq}, regarding the correlation between the emergence of an effective shadow and the presence of a maximum in the angular velocity. Additionally, it is yet unclear how the existence of disconnected stable regions for TCOs might impact the accretion flow, and whether an emission profile resembling that of Eq.~\eqref{eq_luminosity2} could be obtained in practice. Work on such simulations for PSs is underway and will be reported elsewhere.


\section*{Acknowledgments}

This work is supported by the Center for Research and Development in Mathematics and Applications (CIDMA) through the Portuguese Foundation for Science and Technology Funda\c c\~ao para a Ci\^encia e a Tecnologia), UIDB/04106/2020, UIDP/04106/2020,
https://doi.org/10.54499/UIDB/04106/2020 
and https://doi.org/10.54499/UIDP/04106/2020.
The authors acknowledge support from the projects\\
http://doi.org/10.54499/PTDC/FISAST/3041/2020, as well as 
http://doi.org/10.54499/CERN/FIS-PAR/0024/2021 and https://doi.org/10.54499/2022.04560.PTDC.  This work has further been supported by the European Union’s Horizon 2020 research and innovation (RISE) programme H2020-MSCA-RISE-2017 Grant No. FunFiCO-777740 and by the European Horizon Europe staff exchange (SE) programme HORIZON-MSCA2021-SE-01 Grant No. NewFunFiCO101086251. PC is supported by the Individual CEEC program  http://doi.org/10.54499/2020.01411.CEECIND/CP1589/CT0035 of 2020, funded by the FCT.
IS is supported by the FCT grant SFRH/BD/150788/2020 under the IDPASC Doctoral Program. Computations have been performed at the Argus and Blafis cluster at the U. Aveiro.

\bibliographystyle{unsrt}
\bibliography{Refs}

\newpage
\appendix
\label{sec:appendix}
\section{\label{App1}Redshift factor for a ZAMO observer at a finite distance}

This appendix discusses how to compute the redshift factor $\nu_0/\nu$ between the frames $\mathcal{D}$ and $\mathcal{O}$ (see Section~\ref{sec:redshift}). The ratio of the photon frequencies can be related to the ratio of the energies via Planck's constant $h$:

\begin{align}
	\frac{\nu_0}{\nu} =\frac{h \nu_0 }{h \nu} = \frac{E_0}{E}\, ,
\end{align}

\noindent where $E_0$ stands for the photon energy measured in the local emitting frame $\mathcal{D}$ of the accretion disk, and $E$ stands for the photon energy measured in the observation frame $\mathcal{O}$.\\

Similarly to~\cite{Cunha:2016bpi}, $\mathcal{O}$ will be considered as a ZAMO observation frame $\{ \hat{e}_{\left(t \right)}, \hat{e}_{\left(r \right)},  \hat{e}_{\left(\theta \right)},  \hat{e}_{\left(\phi \right)} \}$, which can be expanded in the coordinate basis  $\{\partial_t,\partial_r,\partial_{\theta},\partial_{\phi}\}$ as:

\begin{align} \label{eq2}
	\begin{cases}
		&\hat{e}_{\left(\theta \right)} =A^{\theta} \partial_{\theta} \quad, \quad 	\hat{e}_{\left(r \right)} =A^r \partial_r \\
		&\hat{e}_{\left(\phi \right)} =A^{\phi}\partial_{\phi} \quad ,  \quad	\hat{e}_{\left(t \right)} = \zeta \partial_t + \gamma \partial_{\phi} \\
	\end{cases}
\end{align}

By imposing a Minkowski-type normalization $\hat{e}_{\left(i\right)} \cdot \hat{e}_{\left(j\right)} = \eta_{ij }$, where $\eta_{ij}$ is the flat Minkowski metric, and also by requiring that we recover a standard static frame at spatial infinity, we obtain:

\begin{align} \label{eq:norm}
	&A^{\theta}=\frac{1}{\sqrt{g_{\theta \theta}}}, \quad 	A^{r}=\frac{1}{\sqrt{g_{rr, }}}, \quad 	A^{\phi}=\frac{1}{\sqrt{g_{\phi \phi}}} .\\
        & \nonumber \\
	&\gamma = -\frac{g_{t\phi}}{g_{\phi \phi}} \sqrt{\frac{g_{\phi \phi}}{g_{t \phi}^2 -g_{tt} g_{\phi\phi}}} \, , \quad \zeta=\sqrt{\frac{g_{\phi \phi}}{g_{t \phi}^2 -g_{tt} g_{\phi \phi}}} \, .
\end{align}

In the disk rest frame $\mathcal{D}$, the photon has an energy $E_0$ which is the projection of the photon's 4-momentum $p_\mu$ onto the $\mathcal{D}$ frame's 4-velocity $u^\mu$:

\begin{align}
	E_0=-\left(p_t u^t +p_{\phi} u^{\phi}  \right) \, .
\end{align}

Defining $\eta\equiv -p_\phi/p_t$, and since $u^{\phi} = \Omega u^t$, we can write, 

\begin{align}
	E_0= -p_t u^t \left( 1 - \Omega \eta \right) \, .
\end{align}

The components $p_t$ and $p_{\phi}$ are constants of motion for null geodesics. In the asymptotic limit, $-p_t$ and $p_\phi$ can be interpreted respectively as the photon's energy and axial angular momentum with respect to far-away observers. The ratio $\eta$ can then be regarded as an impact parameter for the light ray.

In addition, the term $u^t$ can be written in a more explicit way using the norm of the 4-velocity $u^{\mu}u_{\mu}=-1$:

\begin{align}\label{eq_rd0}
	u^t= \left(-g_{tt} -2 \Omega g_{t \phi} -\Omega^2 g_{\phi \phi}  \right)^{-1/2}\, .
\end{align}

For a ZAMO observer $\mathcal{O}$ (placed at a finite position), the energy locally measured by the observer can be written as:

\begin{equation} \label{eq_E}
	E=-\left(\hat{e}^{\mu}_{\left( t\right)} p_{\mu}\right)=-\left(\zeta p_t + \gamma p_{\phi} \right) =  -p_t \left(\zeta -\gamma \eta\right)\\
\end{equation}

Combining the previous relations, we obtain an expression for the redshift factor for a ZAMO observer located at an arbitrary position, displayed in the main text:

\begin{align}
	\frac{\nu_0}{\nu}=\left(\frac{1}{\zeta - \gamma \eta} \right)\Bigg|_{\mathcal{O}}  \left(1-\Omega \eta\right) \left(-g_{tt} -2 \Omega g_{t \phi} -\Omega^2 g_{\phi \phi}  \right)^{-1/2}
\end{align}

\end{document}